\documentclass[
 aip,
 amsmath,amssymb,
 reprint,%
 nofootinbib
]{revtex4-1}

\usepackage{graphicx}% Include figure filesj
\usepackage{dcolumn}% Align table columns on decimal point
\usepackage{bm}% bold math
\usepackage{enumerate}
\usepackage{setspace,color,url}
\allowdisplaybreaks[1]

\renewcommand{\mod}[1]{{#1}}

\newcommand{\kt}{k_{\text{B}}T}
\newcommand{\UU}{{\cal U}}
\newcommand{\cc}{\overline c}

\usepackage[dvipsnames]{xcolor}

\usepackage{graphicx}% Include figure files
\usepackage{soul}
\usepackage{mathrsfs}

\newcommand{\lambdamR}{\lambda_{-}^R}
\newcommand{\lambdamL}{\lambda_{-}^L}
\newcommand{\lambdapR}{\lambda_{+}^R}
\newcommand{\lambdapL}{\lambda_{+}^L}

\begin{document}

%\preprint{}
%\preprint{Submitted to The Journal of Chemical Physics}

%%%%%%%%%%%%%%%%%%%%%%%%%%%%%%%%%%%%%%%%%%%%%%%%%%%
%\title{A guide for active selective transport through nanopores}
%\title{Active sieving across nanopores for tunable selectivity}
%\title{Active sieving across forced nanopores for tunable selectivity}

\title{Resonant osmosis across active switchable membranes}
%\title{Frequency dependent osmosis across active membranes}
%\title{Osmotic pumps across active membranes}

%%%%%%%%%%%%%%%%%%%%%%%%%%%%%%%%%%%%%%%%%%%%%%%%%%%
\author{Sophie Marbach}
\affiliation{ Courant Institute of Mathematical Sciences, New York University, New York, New York, USA}
\affiliation{Laboratoire de Physique de l'\'Ecole Normale Sup\'erieure, ENS, Universit\'e PSL, CNRS, Sorbonne Universit\'e, Universit\'e Paris-Diderot, Sorbonne Paris Cit\'e, Paris, France}
\author{Nikita Kavokine}
\author{Lyd\'eric Bocquet}

%\affiliation{Laboratoire de Physique, Ecole Normale Sup\'erieure and CNRS, PSL Research University, 24 rue Lhomond, 75005 Paris, France}
\affiliation{Laboratoire de Physique de l'\'Ecole Normale Sup\'erieure, ENS, Universit\'e PSL, CNRS, Sorbonne Universit\'e, Universit\'e Paris-Diderot, Sorbonne Paris Cit\'e, Paris, France}
\email{lyderic.bocquet@ens.fr}
\date{\today}
%%%%%%%%%%%%%%%%%%%%%%%%%%%%%%%%%%%%%%%%%%%%%%%%%%%
\begin{abstract}
To overcome the traditional paradigm of filtration, where separation is essentially performed upon steric sieving principles, we explore the concept of dynamic osmosis through active membranes. A partially permeable membrane presents a time-tuneable feature that changes the effective pore interaction with the solute and thus actively changes permeability with time. In general we find that slow flickering frequencies effectively decrease the osmotic pressure, and large flickering frequencies do not change it. In the presence of an asymmetric membrane, we find a resonant frequency where pumping of solute is performed and can be analyzed in terms of ratchet transport. We discuss and highlight the properties of this resonant osmotic transport. Furthermore, we show that dynamic osmosis allows to pump solute at the nanoscale using less energy than reverse osmosis. This opens new possibilities to build advanced filtration devices and design artificial ionic machinery. 
\end{abstract} % 600 characters length limit //  current = 601
%%%%%%%%%%%%%%%%%%%%%%%%%%%%%%%%%%%%%%%%%%%%%%%%%%%
%\pacs{ 05.40.-a, 05.60.Cd, 82.39.Wj}
\maketitle

\section{Introduction}

Modern processes for filtration are based on passive sieving principles: a membrane with specific pore properties allows to separate the permeating components from the retentate \cite{marbach2019osmosis}. The domain has been boosted over the last two decades by the possibilities offered by nanoscale materials, such as graphene based or advanced membranes.~\cite{Karnik2014,Geim2014,Siria2013,Picallo2013,Feng2016,tunuguntla2017enhanced,esfandiar2017size} Selectivity requires small and properly decorated pores at the scale of the targeted molecules, and this inevitably impedes the flux and transport, making separation processes costly in terms of energy. Furthermore, standard membranes suffer from an intrinsic limitation: to increase permeability, one must typically increase the size of the pores at the expense of inevitably diminishing selectivity. This is commonly referred to as the selectivity-permeability trade-off.~\cite{Elimelech2016}

\begin{figure}[t]
\includegraphics*[width=0.99\columnwidth]{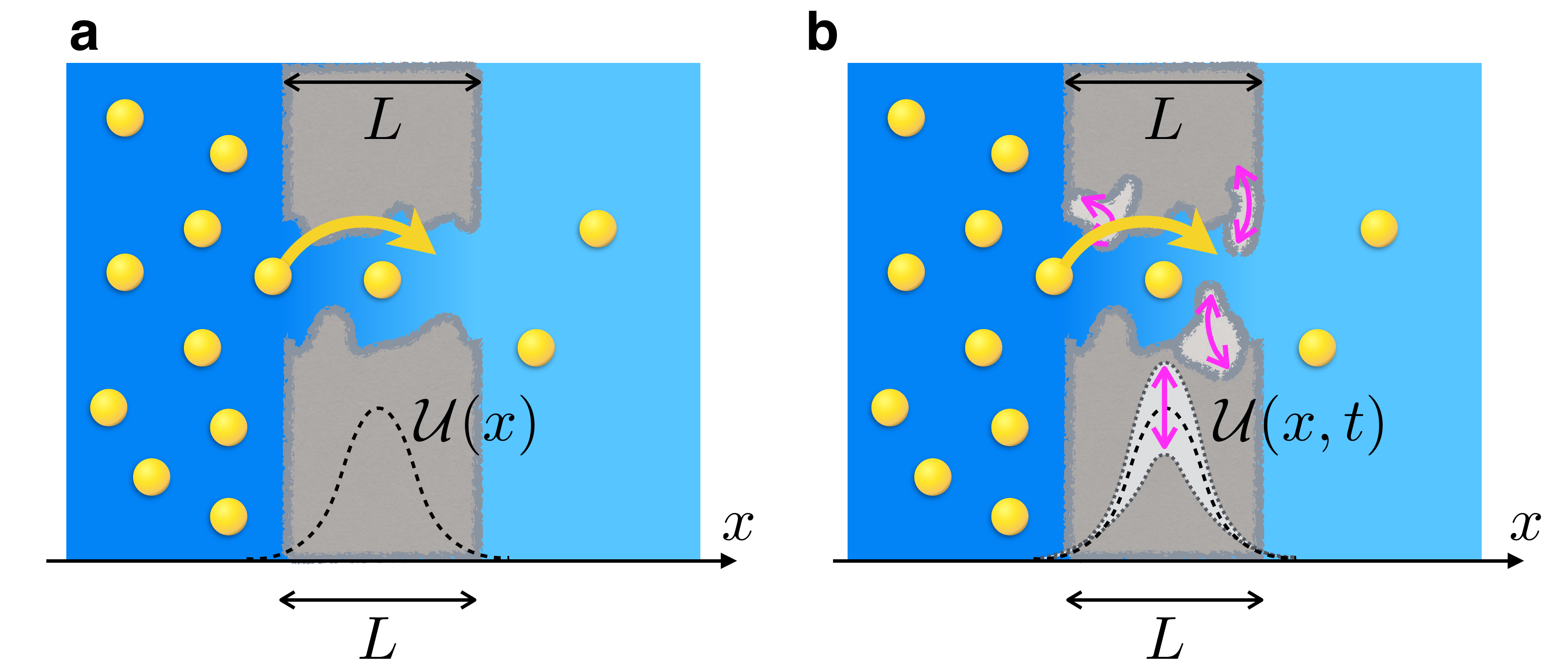}%
\caption{\label{fig:geo} \textbf{Membranes as potential barriers}. (a) Porous membrane seen as an energy barrier $\UU(x)$; (b) The porous membrane has some temporal dependence (for instance time-dependent porous aperture), and can be seen as a time dependent energy barrier $\UU(x,t)$.}
\end{figure}

However, this classical paradigm only considers membranes with fixed properties and pore size, 
%The classical paradigm of these membranes is that they are static in time, 
and therefore the constraints of selectivity-permeability are defined with \textit{static} systems. 
Interestingly, Nature encompasses a number of highly selective and highly permeable porins that operate far from equilibrium, and involve {\it active} parts.~\cite{marbach2019osmosis,wei2016protein,bhabha2011dynamic,Allen2004} Pore shape agitation was identified in some cases to be tightly connected to selectivity properties.~\cite{Noskov2004} Therefore it is natural to revisit the trade-off paradigm by investigating how it is possible to harness non-equilibrium dynamics and active membranes to separate solutes across \textit{active} nanopores,~\cite{marbach2017active,marbach2018transport} see \textit{e.g.} Fig.~\ref{fig:geo}.  
There is accordingly an interesting analogy with active matter 
%The description of active fluids has required to introduce a new thermodynamic framework to account for a wide range of effects.
~\cite{ginot2015nonequilibrium,solon2015pressure,solon2018generalized} and the osmotic pressure generated by active fluids in the vicinity of passive semi-permeable membranes has also been explored.~ \cite{lion2014osmosis,rodenburg2017van} 
However how membrane dynamics may affect osmotic pressure remains to be investigated. 
%Surprisingly, the question of how osmotic pressure is affected by membrane dynamics is still open today and lacks intuitive behavioral rules. 
In this context we explore the concept of {\it  dynamic osmosis} and the possibility of tuning the osmotic pressure via the membrane dynamics. This corresponds to a non-equilibrium situation, which could allow to some extent to bypass the equilibrium constraints of separation. 

Considering a nanopore with some dynamic feature (\textit{e.g.} a flickering aperture, a time-dependent surface charge...), we raise the following questions: how is the osmotic pressure expressed? How does the osmotic pressure depend on the typical timescale of the dynamic feature? 
%We clarify these questions using here a simplified model of a membrane with a kinetic approach.
To adress these question, we will consider a simple, yet insightful kinetic model of membrane separation, in which 
the membrane pores are assimilated to a potential energy barrier $\UU(x)$ across the membrane (see Fig.~\ref{fig:geo}-a) ~\cite{manning1968binary,picallo2013nanofluidic,marbach2017osmotic}. This energy profile is allowed to vary with a typical time scale, modeling the dynamic feature of the membrane (see Fig.~\ref{fig:geo}-b). We show that the osmotic pressure response is a highly non-trivial function of the frequency of the pore oscillations. In specific regimes where the energy barrier is asymmetric,  the osmotic pressure exhibits a resonance
at a characteristic frequency. Interestingly, we will harness %dynamical osmosis harness 
%it is possible to transfer 
the know-how of transport and pumping through oscillating ratchet potentials %to predict the properties dynamical osmosis
~\cite{rousselet1994directional} to predict the properties of dynamical osmosis. This allows in particular to identify the design rules for a minimal osmotic pump. The properties of such active membranes are therefore extremely broad and could be harvested for advanced nanofiltration. Finally, we show that dynamic osmotic solute pumping is energetically less costly than standard reverse osmosis. 

\section{Active nanopore}

%\subsection{Towards a kinetic dynamical membrane model}

%Traditionally, osmotic transport is described as occurring across a semi-permeable membrane, {\it i.e.}, a membrane impermeable to the solute but permeable to the solvent, often water. If two reservoirs with different solute concentrations are put in contact via a semi-permeable membrane, an osmotic pressure builds up between the compartments. This pressure drop is the driving force for a flux of water from the low concentration reservoir 
%to the highly concentrated one, until the thermodynamic equilibrium is reached. For low solute concentrations, the osmotic pressure is expressed by the van\;'t Hoff law, 
%$\Delta \Pi = \kt \Delta c, $
%where $\Delta c$ is the difference in solute concentration between the two reservoirs.~\cite{kedem1961physical}
%The van\;'t Hoff law is derived by equating the solvent chemical potential of the solvent across the membrane.~\cite{gibbs1897semi,guggenheim1985thermodynamics,klotz2008chemical} The osmotic pressure is accordingly defined in terms of equilibrium thermodynamic properties of the system.
%%, well defined in terms of the entropy of mixing between solute and solvent.}
%The membrane characteristics do not appear in this thermodynamic expression for the osmotic pressure.

%\revb{do we need a more detailed intro ?... van't hoff, link to kedem katchalsky and energy barrier model ?}

\subsection{Active membrane model and qualitative considerations}

We consider a porous membrane separating two sub-volumes, containing a solvent and a solute. There is a solute concentration difference between the two sub-volumes, $\Delta C=C_2-C_1$. Following Ref.~\citenum{manning1968binary}, we consider 
a model in which the pores are replaced by a potential barrier (see Fig.~\ref{fig:geo}). Specifically, we model the membrane as an external potential $\UU (x,t)$ acting on the solute {\it only}, and not on the solvent molecules. The membrane thus still remains permeable to the solvent, with a permeance $\mathcal{L}_\text{hyd}$, relating the flux $Q$ to the pressure drop $\Delta p$ in the absence of a concentration difference: $Q=-\mathcal{L}_\text{hyd} \Delta p$.
The potential $\UU (x,t)$ varies only along the $x$ axis across the membrane. 
We denote $L$ the characteristic thickness of the membrane, so that $\UU$ vanishes outside the lateral range $L$, see Fig.~\ref{fig:geo}. \mod{The potential $\UU$ represents any kind of interaction between the solute and the membrane. These could be steric interactions for \textit{e.g.} colloids or large molecules, or electrostatic interactions between charged solutes and a charged membrane, \textit{etc.}} \mod{Although here we consider that the primary interaction is between the solute and the membrane, our model could be extended further to account for specific interactions between the solute and the solvent. To give rise to osmosis -- which is our interest here --, the necessary condition is that solute and solvent do not interact in the same way with the membrane~\cite{marbach2019osmosis}, and therefore to simplify we only consider one interaction.}

\begin{figure}[h!]
\includegraphics*[width=0.99\columnwidth]{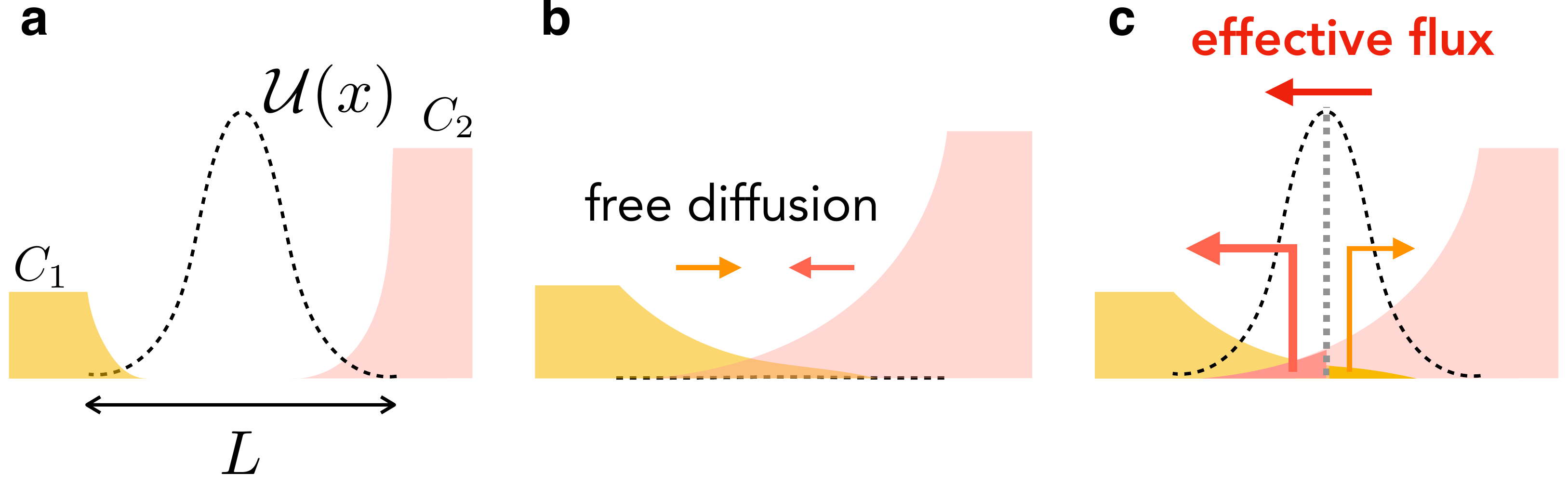}%
\caption{\label{fig:intuition} \textbf{Expected behavior with oscillating symmetric energy barrier.} (a) The energy barrier is initially fully expressed, with the same solute on each side but in different concentrations; (b) when the barrier is decreased, solute can diffuse, eventually mixing up between both sides; (c) when the barrier goes up again,  solute is pushed back outwards, and solute effectively originating from the right hand side ends up on the left hand side, inducing an effective flux of solute. The effective flux is expected to depend on the spatial and temporal characteristics of the potential.}
\end{figure}

If the potential $\UU(x,t)$ is static in time, the above kinetic framework allows to recover, for instance, the van 't Hoff law for osmosis~\cite{manning1968binary,marbach2017osmotic}. Here we are interested in the dynamical case, where a time-dependent pore permeability is modeled by an oscillating potential
\begin{equation}
{\cal U}(x,t)={\cal U}_0\,\phi(x)\left( 1 + \epsilon \,\cos \omega t\right).
\label{uxt}
\end{equation}
The resulting configuration is schematically depicted in Fig.~\ref{fig:intuition}. As the energy profile goes down with time, the concentration profile is accordingly modified, as diffusion brings solute into the membrane. When the energy profile goes up again, solutes diffuse outwards, and the concentration profile flows away accordingly. \mod{A typical system representing such an active membrane could consist in electrically gated pores~\cite{kim2019stacked,kavokine2019ionic} or in mechanically driven pores with some external excitation~\cite{marbach2018transport}. Note as well that nearly every biological nanochannel works in such nonequilibrium conditions with \textit{e.g.} electrical or mechanical gating~\cite{yellen2002voltage}.}

%We expect that the resulting flux (see Fig.~\ref{fig:intuition}), and therefore the osmotic pressure, will be greatly dependent on the timescale of the oscillation $\omega$, but also on the strength $u_0$ of the energy barrier. 

At this stage one can note that the ingredients entering our system are very similar to those composing an oscillating potential ratchet.~
\cite{magnasco1993forced,rousselet1994directional,astumian1994fluctuation,reimann1995thermally,reimann1996brownian,reimann2002introduction,reguera2012entropic}  Therefore, we may expect the flux of solute particles to be strongly dependent on the frequency of forcing, as well as the height $\UU_0$ of the energy barrier. Here we are especially interested in the consequences on the osmotic pressure, for which there is little intuition and no analytic result. 

In the following subsections we give details on how to compute the concentration profile, the effective flux and the osmotic pressure in this oscillating case. In the following steps, we will perform an expansion in $\epsilon$ for any potential shape in order to obtain general results for the osmotic pressure as a function of the frequency. Then in the next sections, we will apply these results
to specific shapes of the potential and obtain explicit results. 

\subsection{Expansion of the Smoluchowski equation} 

In the 1D geometry described above, the solute concentration $c(x,t)$ obeys the time-dependent Smoluchowski equation
\begin{align}
{\partial_t} c =& -\partial_x\left( -D \partial_x c + \lambda c \,(-\partial_x \UU) + v c \right),
\label{Smolu}
\end{align}
where $D$ is the diffusion coefficient and  $\lambda=D/\kt$ the mobility, with $k_{\mathrm{B}}$ and $T$ being the Boltzmann constant and the temperature, respectively.
We further assume a low P\'eclet number limit, $\mathrm{Pe}=v L/D \ll 1$, such that the convective term of Eq.~\eqref{Smolu} is negligible. This is valid for low permeability (nanoporous) membranes \mod{-- note that convective terms are also of higher order in the concentration profile as the velocity field $v$ typically scales as $c \,(-\partial_x \UU)$ from the Stokes for flow eq.~\ref{St}}.

The boundary conditions for the concentration are
\begin{align}
c(x_1,t)=& \,C_1  \nonumber\\
c(x_2,t)=& \,C_2
\end{align} 
with $x_1=-{L}\times\delta_0$ and $x_2=+{L}\times (1 - \delta_0)$ and $\delta_0$ a dimensionless parameter (see \textit{e.g.} Fig.~\ref{fig:pumpingSchematic}). In the following we will also use $\delta_1 = 1 - \delta_0$. 

Using now the expression of $\UU (x,t)$ in Eq.~\eqref{uxt}, the Smoluchowski equation becomes
\begin{align}
{\partial_t} c =& -D\partial_x\left( - \partial_x c +  c {\UU_0\over k_BT}(1+\epsilon \cos \omega t)\,(-\partial_x \phi) \right).
\label{Smolu2}
\end{align}
We expand the solution as
\begin{align}
&c(x,t)= \cc(x) + \epsilon\, \delta c_1(x,t) +\epsilon^2 \delta c_2(x,t) \nonumber \\
&{\rm with}\,\,\, \delta c_1(x,t)={\cal R}e\left[ \delta c_1(x) e^{j\omega t}\right]\nonumber \\
&{\rm and}\,\,\, \delta c_2(x,t)={\cal R}e\left[\delta c_2^0(x)+ \delta c_2^1(x) e^{2j\omega t}\right]
\label{defc}
\end{align}
where $ {\cal R}e$ stands for the real part. This expansion is thought as an expansion in $\epsilon$.
The term $c\times \cos\omega t$ in Eq.~(\ref{Smolu2}) leads at second order to the expected modes in $0$
and $2\omega$, yielding the two second order terms in Eq.~(\ref{defc}). In average over time, we expect the first order terms to vanish. To compute relevant quantities such as the osmotic pressure and the average flux through the membrane, we thus need to perform the expansion up to second order. 

\subsection{Concentration profile equations}

In this section we present an analytic derivation for the concentration profile and solute flux up to second order. For readability we nondimensionalize the equations using $\tilde x=x/L$, $\tilde t= t/\tau_0$, $\tilde{c} = c/C_0$ (where $C_0 = (C_1+C_2)/2$, and $\tilde \omega = \omega/\omega_0$, with $\omega_0^{-1} = \tau_0 = L^2/D$, 
$u_0={\UU_0\over k_BT}$). We then drop the tilde signs to simplify. We also write $\Delta c = {(C_2 - C_1)/}{C_0}$, and fluxes are nondimensionalized by $D\, C_0/L$.

\subsubsection{Zeroth order equation}
The zeroth order solution $\cc(x)$ is assumed to be stationary and thus obeys
\begin{align}
0=& -\partial_x\left( - \partial_x \cc +  \cc\, {u_0}\,(-\partial_x \phi) \right),
\label{Smolu1}
\end{align}
whose solution is (see Ref.~\citenum{marbach2017osmotic} for details)
\begin{align}
& \cc (x)=e^{- u_0 \phi(x)}\nonumber \\
&-(\Delta c)\, e^{-u_0 \phi(x)} { \int_{x}^{\delta_1} dx^\prime\, \exp[+ u_0 \phi(x^\prime)]\over \int_{-\delta_0}^{\delta_1} dx^\prime\, \exp[+u_0 \phi(x^\prime)]},
\label{profile}
\end{align}
%
%where $\beta=1/\kt$.
%{\color{purple} Expression for  flux $J_0$.}
The corresponding flux to zeroth order is
\begin{equation}
J_0=-(1-\sigma_0) \times {\Delta c}
\label{J0}
\end{equation}
with the rejection coefficient $\sigma_0$ defined as
\begin{equation}
\sigma_0= 1-{1\over \int_{-\delta_0}^{\delta_1} dx^\prime\, \exp[+u_0\phi(x^\prime)]}.
\label{sigma0}
\end{equation}

\subsubsection{First order equation}
The equation for the time-dependent concentration $\delta c(x,t)$ at order 1 is
\begin{align}
\partial_t \delta c_1(x,t)=&  \partial_{xx} \delta c_1(x,t)   - {u_0}\,\partial_x [ \delta c_1(x)  (-\partial_x \phi) ]\nonumber\\
&-{u_0}\cos(\omega t)\,\partial_x [\cc(x)  (-\partial_x \phi)] .
\label{Smolu2b}
\end{align}
Accordingly, the first order complex amplitude,  $\delta c_1(x,t)={\cal R}e\left[ \delta c_1(x) e^{j\omega t}\right]$ (see  Eq.~(\ref{defc})), obeys
%\begin{align}
%j\Omega \delta c(x)=& D \partial_{xx} \delta c(x)   - D\partial_x \delta c(x)  {\UU_0\over k_BT}\,(-\partial_x \phi) \nonumber\\
%&-D\partial_x \cc(x) \epsilon \,{\UU_0\over k_BT}\,(-\partial_x \phi) ,
%\label{Smolu3}
%\end{align}
%
%The above equation then rewrites
\begin{align}
j\omega \delta c_1(x)=& \partial_{xx} \delta c_1(x)   - {u_0}\,\partial_x [\delta c_1(x)  (-\partial_x \phi)] \nonumber\\
&-{u_0}\,\partial_x [\cc(x) (-\partial_x \phi)] ,
\label{Smolu4}
\end{align}
The boundary conditions are assumed to be $\delta c(x= -\delta_0)=\delta c(x= \delta_1)=0$. The last term of Eq.~(\ref{Smolu4}) is a driving term. 

This equation can be solved for some specific forms of $\phi(x)$, and we come back to analytic solutions in the following sections. In the end we will find $\delta c_1 (x,t) =  \delta c_1 (x)  \cos ( \omega t + \varphi)$ where the phase $\varphi$ depends on all parameters. %and is crucial. 

\subsubsection{Second order equation}

As pointed out above, the second order is a sum of zero frequency and $2\omega$ terms: $\delta c_2(x,t)={\cal R}e\left[\delta c_2^0(x)+ \delta c_2^1(x) e^{2j\omega t}\right]$. We focus on the zero
frequency term, $\delta c_2^0(x)$, which is relevant for the flux and osmotic pressure, while the $2\omega$ term
will not contribute and averages to zero.

The second order static term obeys the equation
\begin{align}
0 =& \partial_{xx} \delta c_2^0(x) - u_0\,\partial_x[\delta c_2^0(x) (-\partial_x \phi)]  \\
&-  {u_0\over 2}
\partial_x[ |\delta c_1(x)| (-\partial_x \phi)] \cos (\varphi)
\label{Smolu20}
\end{align}
where the last term originates from the time average of the first order term over one period. One can also just solve Eq.~(\ref{Smolu20}) in the complex domain and we do that in the following.
We assume the following boundary conditions: $\delta c_2^0(x= -\delta_0)=\delta c_2^0(x= \delta_1)=0$.

Eq.~(\ref{Smolu20}) can be easily solved. Defining the second order flux as
\begin{equation}
J_2=-\partial_x  \delta c_2^0(x)+ u_0 (-\partial_x \phi) \delta c_2^0(x) +{u_0\over 2}
\partial_x[ \delta c_1(x) (-\partial_x \phi)],
\end{equation}
one has $J_2={\rm const}$. This yields
\begin{equation}
J_2={1\over 2} {\int_{-\delta_0}^{\delta_1} dx\,  \delta c_1(x) (-\partial_x \exp[u_0\phi])
\over \int_{-\delta_0}^{\delta_1} dx\, \exp[u_0\phi]}
\label{J2}
\end{equation}
and 
\begin{align}
 \delta c_2^0(x)=&-J_2 e^{-u_0 \phi(x)} \int_{-\delta_0}^{x} dx^\prime\,e^{u_0 \phi(x^\prime)} \nonumber \\
&+{e^{-u_0 \phi(x)}\over 2}  \int_{-\delta_0}^{x} dx^\prime\,\delta c_1(x)\, (-\partial_x \exp[u_0\phi(x^\prime)]).
\end{align}

\subsection{Dynamic osmotic pressure and flux}

\subsubsection{Osmotic pressure}
We now turn to the expression of the osmotic pressure. We write accordingly the force balance on the fluid (composed of the solvent and the solute). It is crucial to remark that the membrane will act on the fluid as an external force, $-\partial_x \UU$, exerted on the solute molecules. This is due to solute and solvent being in a dense interacting phase, where the force acts on the whole fluid volume as solvent molecules are dragged along the solute. This is expressed writing the force balance on the fluid, represented by the Stokes equation along the $x$ direction (here fully dimensionalized):
\begin{equation}
\rho \partial_t v = -\partial_x p + c(x) (-\partial_x \UU)+ \eta \nabla^2 v,
\label{St}
\end{equation}
where 
$p$ is the fluid pressure, $v$ is the flow velocity of the fluid in the $x$ direction, $\eta$ is the fluid viscosity and $\rho$ its density.
The driving force inducing solvent flow is accordingly written in terms of an apparent pressure drop, $-\partial_x \mathcal{P}=-\partial_x p+c(x)(-\partial_x\UU)$. 
The membrane, via its potential $\UU$, will therefore create a pressure force on the fluid, which writes per unit surface 
\begin{equation}\label{eq_Fc}
\sigma \Delta\Pi=\int_{-\delta_0 L}^{\delta_1 L} dx\, c \,(-\partial_x \UU).
\end{equation}
$\Delta\Pi$ is identified as the osmotic pressure which in the dilute case takes the simple van 't Hoff expression $\Delta\Pi = k_B T \Delta C$; $\sigma$ is a screening parameter that takes into account the specificities of the membrane. \mod{Assuming that the time scale to establish the flow is much faster than the time scale of oscillation of the potential barrier}, the fluid flux will therefore write $Q = - \mathcal{L}_{\rm hyd} (\Delta p - \sigma  \Delta \Pi)$. \mod{At high forcing frequencies, this assumption should be reconsidered to account for inertial effects and may lead to enhanced or decreased behaviors}. 

 Here we are interested in the averaged effective force over a period
$\langle \sigma \Delta\Pi\rangle$.
Following the previous formal expansion $c(x,t)=\cc(x) + \epsilon\delta c_1(x,t) + \epsilon^2 \delta c_2(x,t)$, we expand the osmotic pressure contribution as
\begin{equation}
\Delta \Pi_{\rm app} \equiv  \sigma_{\rm app} \Delta \Pi \equiv \langle \sigma \Delta\Pi\rangle  = \Pi_0 +  \Pi_1 +  \Pi_2
\end{equation}
corresponding to contributions of the zeroth, first and second order terms in the concentration; $\sigma_{\rm app}$ is an apparent screening parameter. Note that both terms $ \Delta \Pi_1$ and $\Delta \Pi_2$ are of order 2 in $\epsilon$. We come back to nondimensionalized equations, where $\Delta \Pi$ is nondimensionalized by $k_B T C_2$. 

\paragraph{To zeroth order}
The corresponding osmotic pressure contribution matches the stationary solution (see also Ref.~\citenum{marbach2017osmotic}) and
writes 
\begin{equation}
\Pi_0= \sigma_0\times \Delta c \nonumber\\
\label{reflectDilute}
\end{equation}
where $\sigma_0$  is defined by Eq.~(\ref{sigma0}). Note that the osmotic pressure contribution at zeroth order satisfies the following relation to the particle flux: $\Pi_0 = J_0 + \Delta c$ (in dimensionless form).  %This relation is well known yet it is not obvious that it should hold at higher orders.

\paragraph{To first order} 
We average the solution over a period to obtain
\begin{equation}
\Pi_1 ={\epsilon^2  \over 2} \int_{-\delta_0}^{\delta_1} dx^\prime\, {\cal R}e\left[\delta c_1(x^\prime)\right] \times u_0 (-\partial_x\phi)(x^\prime)
\end{equation}

\paragraph{To second order}
Only the zero frequency term $\delta c_2^0(x)$ contributes to the osmotic pressure, so that:
\begin{equation}
\Pi_2 = {\epsilon^2} \int_{-\delta_0}^{\delta_1} dx^\prime\, {\cal R}e\left[\delta c_2^0(x^\prime)\right] \times u_0 (-\partial_x\phi)(x^\prime)
\end{equation}

\subsubsection{Relation to the particle flux}

The (fully dimensionalized) solute flux is defined as
\begin{equation}
J=-D\partial_x c + {D\over k_BT} c (-\partial_x \UU).
\end{equation}
From Eq.~(\ref{Smolu}) one then deduces that the time averaged flux  $\langle J \rangle$ obeys $\partial_x \langle J \rangle=0$, so that
\begin{equation}
\begin{split}
\langle J \rangle(x) &= -D\partial_x \langle c(x,t)\rangle  + {D\over k_BT} \langle c(x,t) (-\partial_x \UU)(x,t)\rangle \\
&= {\rm const}
\end{split}
\end{equation}
Using $ \Delta\Pi_{\rm app}= \langle \int_{x_1}^{x_2} dx\, c \,(-\partial_x \UU) \rangle$, one can integrate this result to obtain
\begin{equation}
  \Delta\Pi_{\rm app} =k_BT [C_2-C_1] + {\kt L \over D}  \langle J \rangle
\label{eq:relation} 
 \end{equation}
and in dimensionless form:
\begin{equation}
 \Delta\Pi_{\rm app} = \Delta c +  \langle J \rangle.
 \end{equation}
Therefore the osmotic contribution may be related to the solute flux at any order, and also in out-of-equilibrium conditions. We stress that Eq.~\ref{eq:relation} is highly interesting because it allows, from the description of the solute flow, to quantify the osmotic pressure contribution. In general it is difficult to compute the osmotic pressure contribution directly, and such a symmetry relation is of great help to obtain the expression for the apparent osmotic pressure. 
 
 The averaged flux can be calculated as $\langle J \rangle=J_0+\epsilon^2\,J_2$, with the first order term averaging to zero.
 Using Eqs.~(\ref{J0})-(\ref{J2}), one deduces 
\begin{align}
 \Delta\Pi_{\rm app}=&\sigma_0 \Delta c + (1-\sigma_0)\, \epsilon^2\, \times ... \nonumber \\
  &\int_{-\delta_0}^{\delta_1} dx^\prime\,  {\cal R}e\left[\delta c_1\right](x^\prime)(-\partial_x \exp[u_0\phi(x^\prime)])
  \label{PressionOsmotique}
  \end{align}
  where the rejection coefficient $\sigma_0$ is defined in Eq.~(\ref{sigma0}).
  
  One can check that this expression matches the direct calculation of the osmotic pressure from
  the force, see above. % I did that it's fine
 Writing $ \Delta\Pi_{\rm app} = \sigma_{\rm app} \Delta \Pi$ with  $\Delta \Pi = \kt\,[C_2-C_1]$ (now fully dimensionalized), $\sigma_{\rm app}$  plays the role of an {\it apparent rejection coefficient}. Note that $ \sigma_{\rm app}$ may depend on the concentrations $C_1$ and $C_2$ and simplifies to
  \begin{align}
 \sigma_{\rm app}[\omega,C_1,C_2]&=\sigma_0 + {(1-\sigma_0)\, \epsilon^2\over \Delta c}\, \times ... \nonumber \\
  &\int_{-\delta_0}^{\delta_1} dx^\prime\,  {\cal R}e\left[\delta c_1\right](x^\prime)(-\partial_x \exp[u_0\phi(x^\prime)])
  \label{Rejection}
  \end{align}

\subsection{Explicit solution for the triangular potential}

In the following we will apply these results to the specific case of a triangular shape for the potential $\mathcal{U}(x)$. 
This allows to obtain explicit analytic expressions for the concentration profile as a function of frequency. 
The analytic expressions are however cumbersome and 
%An explicit analytic solution may be found for the apparent reflection coefficient in the case of a triangular potential $\mathcal{U}(x)$. 
we report the derivation and expressions in Appendix A. In the following we will focus on the implications of this analysis.

\section{Symmetric barrier, towards osmosis on demand}

We investigate first the symmetric barrier case (typically as in Fig.~\ref{fig:analyticalSymmetric}-c), using both the analytic results and standard numerical simulations (see Appendix B for numerical simulation details). We explore a range of modulation frequencies and modulation depths $\epsilon$ while keeping the height $u_0$ of the energy barrier fixed. In Fig.~\ref{fig:analyticalSymmetric}-a and b we show the analytic and numerical results for the apparent flux $\langle J \rangle $ and the apparent osmotic pressure $\Delta\Pi_{\rm app}$. The analytic expansion at small $\epsilon$ is in fairly good agreement with the full numerical simulation as long as $\epsilon \lesssim 0.5$.

\begin{figure}[h!]
\includegraphics*[width=0.99\columnwidth]{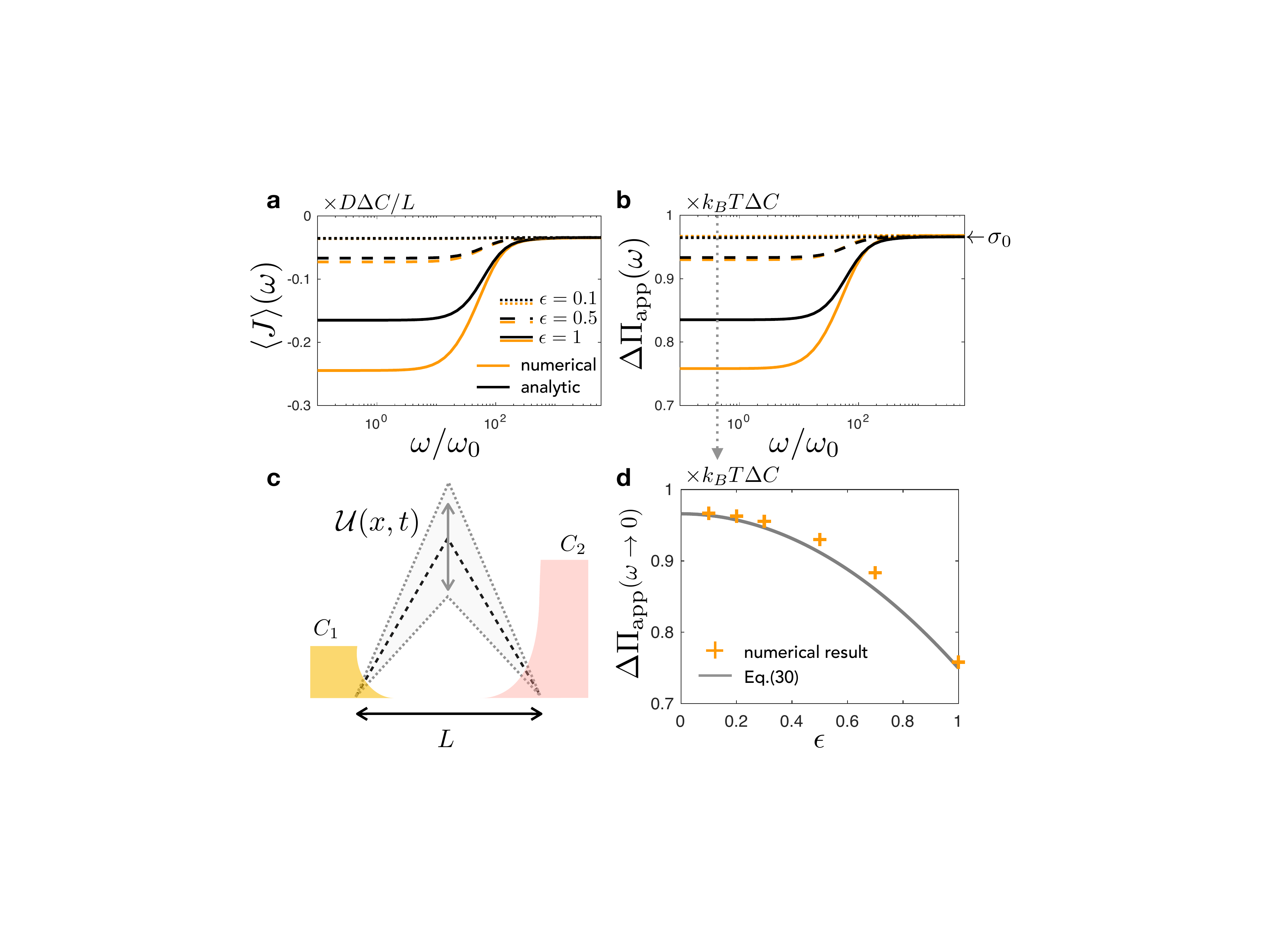}%
\caption{\label{fig:analyticalSymmetric} \textbf{Transport through an active symmetric barrier}. (a) Average solute flux dependence on $\omega/\omega_0$ when $C_2 = 1.82 \, C_0$ and $C_1 = 0.18 \, C_0$ and $U_0 = 5 k_B T$ where $C_0$ is an arbitrary concentration unit. Note that the solute flux is negative corresponding from solute current going from right to left as expected from standard relaxation in sketch (c). It is renormalized by $\Delta C$. (b) Apparent osmotic pressure $\Delta\Pi_{\rm app}$ as a function of $\omega/\omega_0$ for same parameters. The osmotic pressure is given in units of $k_B T \Delta C$. (c) Sketch showing the potential barrier oscillation between two solute reservoirs at different concentrations. (d) Numerical results and prediction from Eq.~(\ref{limitSigma}) for the low frequency apparent osmotic pressure.}
\end{figure}

For large forcing frequencies, the apparent osmotic pressure $\Delta\Pi_{\rm app}$ approaches the usual van\,'t Hoff contribution $k_B T \Delta C$, in other words the rejection coefficient plateaus to a constant value independent of the frequency, as for static membranes.
%converges to a large finite value. For large forcing frequencies, 
%approaches here unity for the . 
In this regime, the concentration profile does not follow the temporal variations of $\UU(x,t)$ and thus effectively sees only its time-averaged value $\langle \UU(x,t) \rangle_t = \UU_0 \phi(x) $. We thus expect:
\begin{equation}
\sigma_{\rm app}(\omega \rightarrow \infty) = \sigma_0
\end{equation}
%This result is coherently independent of 
As expected, this result does not depend on $\epsilon$. It is indicated in Fig.~\ref{fig:analyticalSymmetric}-b by a small horizontal arrow. 

For very low forcing frequencies, we expect the concentration profile at any time $t$ to be in quasi-static equilibrium with the 
%energy profile 
potential, 
%taken at that time point $t$ 
so that:
\begin{equation}
\sigma_{\rm app}(\omega \rightarrow 0) = \left\langle 1 - \frac{1}{\int_{-\delta_0}^{1 - \delta_0} dx' e^{u_0\phi(x') (1 + \epsilon\cos \omega t)}} \right\rangle_t;
\end{equation}
the latter may be approximated at small $\epsilon$ and in the case of a triangular symmetric potential ($\delta_0 = 1/2$) one gets
\begin{equation}
\sigma_{\rm app}(\omega \rightarrow 0) \simeq \sigma_0 - \frac{\epsilon^2}{4}(1-\sigma_0)e^{u_0} u_0^2 \frac{1+e^{u_0}}{(1-e^{u_0})^2}
\label{limitSigma}
\end{equation}
Thus, when $\epsilon$ increases, we expect a decrease of $\sigma_{\rm app}$. That is not necessarily obvious since the barrier effectively goes \textit{up and down} in cycles. 
%In fact this result shows that events where the barrier goes down let solute pass effectively more than solute is blocked when the barrier is raised higher. 
This demonstrates in fact that for a given amount of energy, more solute flux is gained by lowering the barrier by that amount than is lost due to raising the barrier by that same amount.
We plot Eq.~(\ref{limitSigma}) as a function of $\epsilon$ in Fig.~\ref{fig:analyticalSymmetric}-d, and the values obtained with the numerical results. The approximation of Eq.~(\ref{limitSigma}) is very robust in reproducing the numerical results. 

These results show that the osmotic pressure contribution is strongly affected by the active component of the membrane. It is therefore possible to tune the osmotic pressure, and achieve ``on demand" values. Such a rich behavior is achieved while only assuming a symmetric potential profile $\UU(x,t)$. In the following, we seek the osmotic pressure response with an asymmetric potential profile, which is expected to be even more varied, and explore the consequences for filtration and separation.
%Inspired by potential ratchets that may achieve pumping of solute with assymetric potential profiles, we seek the osmotic pressure response for assymetric potential profiles. The osmotic pressure response is expected to be even more varied and we explore consequences for filtration and separation in the following part.
%\begin{figure}[h!]
%\includegraphics*[width=0.99\columnwidth]{Figure-New-4.pdf}%
%\caption{\label{fig:analytical} Apparent rejection coefficient $\sigma_{\rm app}$, when $\UU_0 = 5 k_B T$, $\Delta C = 0.9$. }
%\end{figure}

\section{Asymmetric barrier: osmotic resonance}

\subsection{Towards an osmotic pump and sink}
\begin{figure}[h!]
\includegraphics*[width=0.99\columnwidth]{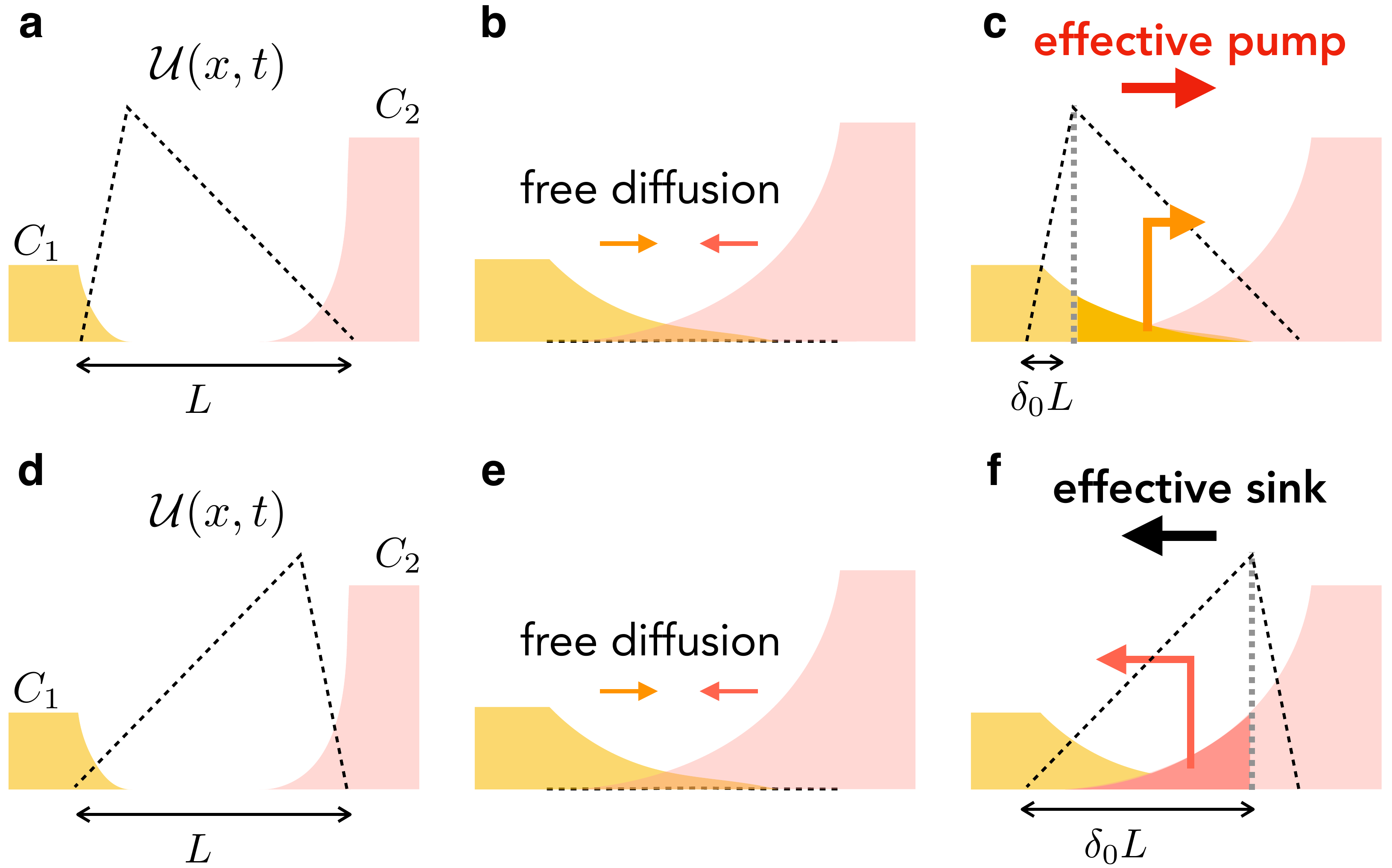}%
\caption{\label{fig:pumpingSchematic} \textbf{Principle for osmotic pump and sink.} (a) Initial configuration of the asymmetric energy barrier $\UU(x,t=0)$ with concentration imbalance; (b) when the barrier is decreased,  the solute diffuses inwards; (c) when the barrier increases back, solute that crossed the maximal point will be flushed towards the right. If the frequency is well adjusted, essentially only solute from the left hand side will have diffused past the barrier and will be flushed to the high concentration reservoir, therefore acting as a pump. The process iterates back to (a). (d) Initial configuration of an effective sink with initial configuration inverted as compared to (a); (e) when the barrier decreases, solute diffuses  inwards; (f) when the barrier increases again at an appropriate time, the solute from the right has diffused beyond the maximal point, and is effectively flushed to the left, thus increasing the effective flux as compared to a symmetric barrier. The process iterates back to (d). This increased diffusion is termed as a 'sink'. }
\end{figure}

In this part we turn to asymmetric potential profiles, and investigate their consequences on osmotic pressure. We are inspired by the classical results on potential ratchets~\cite{rousselet1994directional,magnasco1993forced}. Under an oscillating asymmetric potential profile, one may expect non-trivial pumping of the solute to occur for specific values of the frequency and potential shape. The qualitative principle of this ratchet-type mechanism is sketched in Fig.~(\ref{fig:pumpingSchematic}) for various potential asymmetries, highlighting that an oscillating potential may lead to pumping, or, conversely, accelerate solute diffusion (`sink' regime). Moreover, an oscillating barrier is know to induce the so-called stochastic resonance phenomenon. 
\cite{reimann2002introduction}
Therefore, because of the fundamental relation between the osmotic rejection coefficient and the solute flux demonstrated in Eq.~(\ref{eq:relation}), 
this various effects on the solute flux should convert into a non-trivial resulting osmotic pressure acting on the fluid. The stochastic resonance phenomenon observed on the flux is therefore expected to result in an ``osmotic resonance''. This is what we clarify in the present section.

%a resonance in the solute flux. Depending on the asymmetry, we except either a positive resonance, the oscillating barrier acting as a pump; or a negative resonance, the oscillating barrier acting as a sink. 

%Interestingly, since we are viewing the asymmetric potential here as a model for a semi-permeable membrane, and since we have shown that the fundamental relation between the osmotic rejection coefficient and the solute flux (Eq.~(\ref{eq:relation})) holds out-of-equilibrium, we expect a resonance in the osmotic pressure corresponding to the resonance in solute flux. We explore this osmotic pressure resonance in more detail in the following.

%Inspired by the work on potential ratchets~\cite{rousselet1994directional} we turn to asymmetric potential profiles, and investigate their consequences on osmotic pressure. With an oscillating asymmetric potential profile, we expect a resonance in the flux (either a positive resonance, the oscillating barrier acting as a pump; or a negative resonance, the oscillating barrier acting as a sink), as is recalled in detail in Fig. ~\ref{fig:pumpingSchematic}. From the fundamental relation between the osmotic rejection coefficient and the solute flux eq.~\ref{eq:relation}, we therefore expect a resonance in the osmotic pressure contribution, opening numerous potential avenues for osmotic pressure tuning and control. We explore the consequences on osmotic pressure below in more detail.

\subsection{Characterization of the osmotic resonance, time scales and amplitude}

\subsubsection{Osmotic resonance}
\begin{figure}[h!]
\includegraphics*[width=0.99\columnwidth]{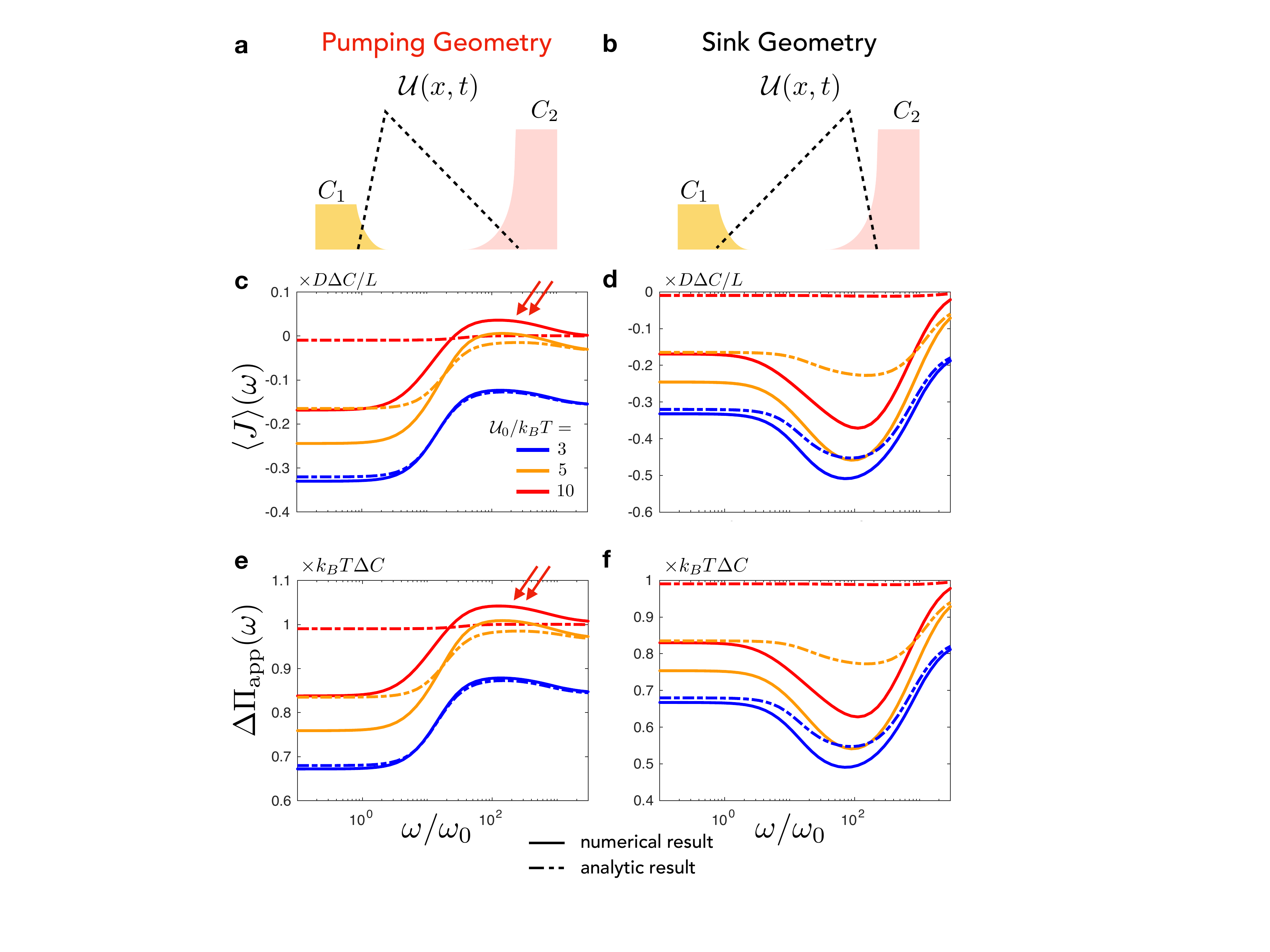}%
\caption{\label{fig:OsmoticPump} \textbf{Osmotic pump and osmotic sink} (a) Schematics showing the geometry relevant for pumping, with a steep energy barrier near the low concentration reservoir, $C_2 > C_1$ and $\delta_0 < 0.5$. (b) Schematics showing the opposite geometry relevant to a sink, with $C_2 > C_1$ but $\delta_0 > 0.5$. (c) and (d) Simulated (solid lines) and analytic (dashed lines) results for the effective normalized flux $\langle J \rangle (\omega)$ and (e) and (f) osmotic pressure $\Delta\Pi_{\rm app}(\omega)$. The results are plotted for several energy barrier strengths in different colors, and $\epsilon = 1.0$ for simplicity. The analytic curve for $\mathcal{U}_0/k_B T = 10$ shows a very small positive flux around the resonance.  In the pumping geometry, $\delta_0 = 0.1$ was taken while in the sink $\delta_0 = 0.9$. For all data $C_2 = 1.82 \, C_0$ and $C_1 = 0.18 \, C_0$.}
\end{figure}

As a proof of principle, we compute the solute flux and apparent osmotic pressure in the case of an asymmetric potential profile. We use both our analytic expansion and standard numerical simulations (see Appendix B). We show the results for the pumping geometry and the sink geometry and for different barrier strengths in Fig.~\ref{fig:OsmoticPump}. Note that the analytic expansion is quite robust but at high energy barrier strengths $\mathcal{U}_0/k_BT$ and at large $\epsilon$ it deviates quantitatively from the simulations (though the observed trends are rather similar). In the case of numerical simulations,  $\Delta\Pi_{\rm app}$ and $\langle J \rangle$ are obtained independently and are in good agreement with the relation of Eq.~(\ref{eq:relation}). 

First, we clearly observe a resonance in both cases in the solute flux and in the apparent osmotic pressure. A pumping regime can indeed be achieved (left panels with $\langle J \rangle>0$ while $C_2>C1$). In terms of osmotic pressure, this translates into an apparent osmotic pressure \textit{greater} than $k_B T \Delta C$ -- or an apparent osmotic reflection coefficient \textit{greater} than 1. This excess osmotic pressure translates into fluid flow. Therefore, if hydrostatic pressure does not equilibrate osmotic pressure, an increased flow of the fluid (including the solvent and solute) is observed in the active osmotic pump regime (in contrast to the static case). 

Second, we clearly observe strong variations of the apparent osmotic osmotic pressure, that eventually can lead to a vanishing or a negative osmotic pressure in the sink geometry in some frequency range; see Fig.~\ref{fig:TimeScaleU0}-b or Fig.~\ref{fig:Alternative}-f. One may therefore tune the sign of the osmotic pressure contribution. When the apparent osmotic pressure is negative, this leads to a flow of fluid against the concentration gradient (towards the dilute side). This fluid flow is accompanied by a flow of solute towards the dilute side. If the permeability of the system is important, one may therefore expect a net pumping of the fluid (hence water). %but the details of the membrane have to be taken into account to fully confirm this assumption.

%\vspace{3mm}

To further illustrate the origin of this phenomenon, it is interesting to investigate a simple toy model with an ON/OFF potential instead of a sinusoidal time dependence. This allows to obtain analytic expression for the frequency dependent osmotic pressure. We report these results in Appendix C. While such results do not aim at a quantitative comparison, they highlight the phenomenon of osmotic resonance in both the pump and sink regime, see Fig.\ref{fig:Alternative}. 

\subsubsection{Resonance frequency}
\begin{figure}[h]
\includegraphics*[width=0.99\columnwidth]{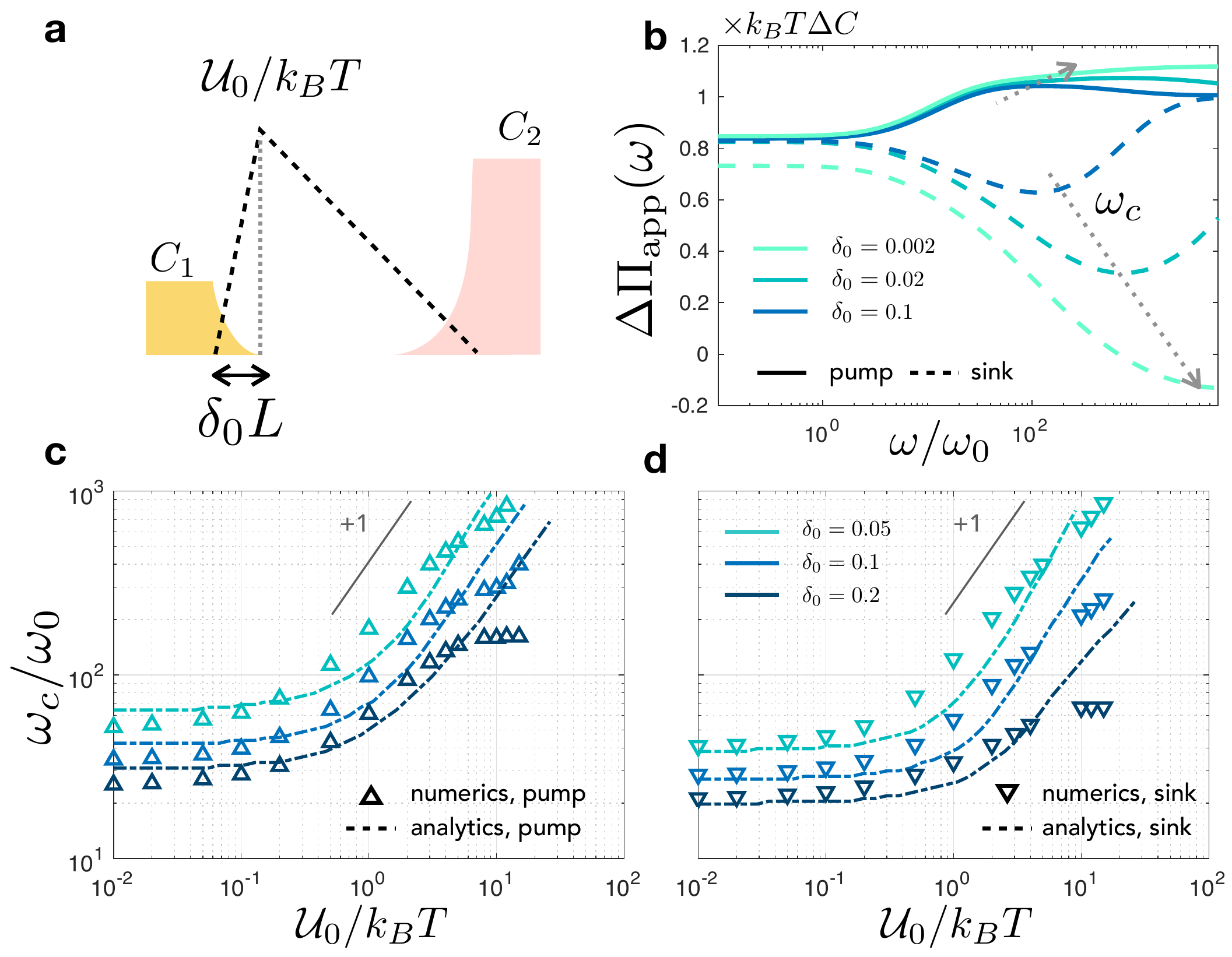}%
\caption{\label{fig:TimeScaleU0} \textbf{Resonance frequency of  active osmosis as a function of the barrier strength $\mathcal{U}_0/k_B T$} (a) Osmotic pump geometry and parameters; (b) Apparent rejection coefficient $\sigma_{\rm app}$ calculated from simulations with respect to the forcing frequency $\omega$ for different values of the asymmetry parameter $\delta_0$, $\UU_0/k_B T = 10$ and $\epsilon = 1.0$. (c) and (d) Resonance frequency $\omega_c$ with respect to the forcing strength $\UU_0/k_B T$ at different $\delta_0$ (same color scale for both graphs) and $\epsilon = 0.5$, for the sink and the pump geometries. Analytical curves are obtained from the expansion discussed in the main text. A scaling law with slope $\omega_c/\omega_0 \propto \UU_0/k_B T$ is indicated in gray. Values for the concentrations are $C_2 = 1.82 C_0$ and $C_1 = 0.18 C_0$ in the pump configuration with $\delta_0<0.5$, and vice-versa for the sink configuration. }
\end{figure}

We now investigate in more detail the resonance frequency $\omega_c$ at which osmotic resonance occurs. It is strongly dependent on the parameters of the system (see for example Fig.~\ref{fig:TimeScaleU0}-b), \textit{e.g.} on the parameters determining the membrane interactions with the solute (barrier strength $\mathcal{U}_0/k_B T$ and asymmetry parameter $\delta_0$).%, and second on the concentration difference (as $\sigma_{\rm app}$ may well depend on the concentration parameters).  actually this is all linear

%\paragraph{Resonance dependence on the potential geometry}
In the pump or the sink process, there are two time scales of interest: \mod{(i) a diffusive time scale that describes the typical time that the solute takes to reach the maximal barrier point (when the barrier is down) and (ii) an advection time scale corresponding to the time it takes to ``slide down" to the other side when the barrier is up again.}. \mod{Let us take the example of the sink process to evaluate these time scales. For the sink process the diffusive time scale writes 
\begin{equation}
\tau_{\rm diff} = \frac{L^2\delta_0^2}{2D}
\end{equation}
as $\delta_0$ is the distance between the highly concentrated side and the barrier peak.} \mod{The advection process corresponds to sliding down the other side of the barrier. It thus takes place with a velocity that is the mobility multiplied by the force $\frac{D}{k_B T} \partial_x \mathcal{U} = \frac{D}{k_BT} \frac{U_0}{L(1-\delta_0)}$. Therefore the advection time scale writes }
\begin{equation}
\tau_{\rm adv} = \frac{L^2(1-\delta_0)^2}{D} \frac{k_BT}{U_0}. 
\end{equation}
%For the process to work in an optimal way, we 
At the resonance, one expects the period of oscillation of the barrier  to be equal to the maximal time scale for the pump or sink process, so that $\tau_c = \mathrm{max} \left( \tau_{\rm diff}, \tau_{\rm adv} \right)$, and therefore the resonance frequency obeys \mod{
\begin{equation}
\omega_c^{\rm sink}/\omega_0 \sim \mathrm{min} \left( \frac{1}{\delta_0^2}, \frac{U_0}{k_B T} \frac{1}{(1-\delta_0)^2}\right) 
\label{eq:omegac}
\end{equation}
and similarly
\begin{equation}
\omega_c^{\rm pump}/\omega_0 \sim \mathrm{min} \left( \frac{1}{(1 -\delta_0)^2}, \frac{U_0}{k_B T} \frac{1}{\delta_0^2}\right).
\label{eq:omegacpump}
\end{equation}}

\begin{figure}[h]
\includegraphics*[width=0.99\columnwidth]{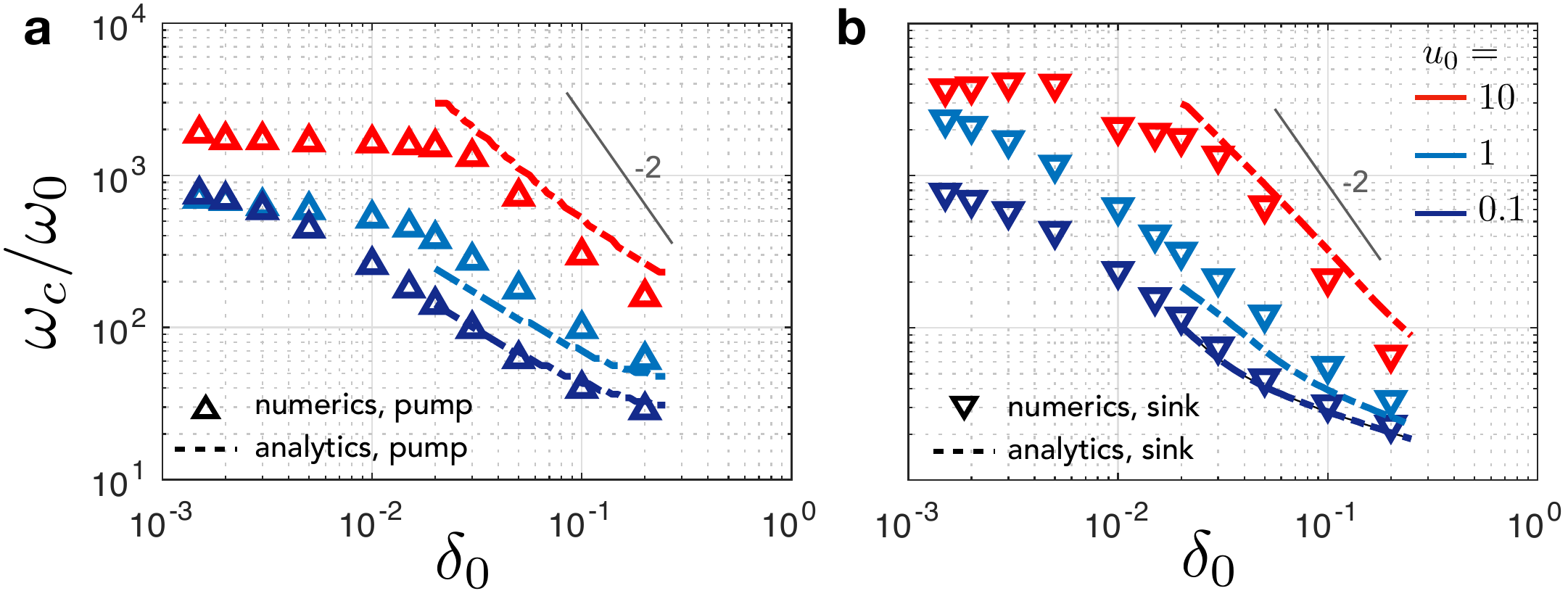}%
\caption{\label{fig:TimeScaled0} \textbf{Resonance frequency of the active osmotic barrier with respect to its asymmetry $\delta_0$}: (a) pumping configuration with $C_1<C_2$ (here  $C_2 = 1.82 C_0$ and $C_1 = 0.18 C_0$); and (b) sink configuration with $C_1>C_2$ (here  $C_1 = 1.82 C_0$ and $C_2 = 0.18 C_0$).
In both panels, the resonance frequency $\omega_c$ is plotted with respect to the asymmetry parameter $\delta_0$, at different forcing strengths $u_0 = \UU_0/k_B T$ (same color scale for both graphs), for the sink and the pump geometries. Numerical and analytical data are for $\epsilon = 0.5$. A scaling law with slope $\omega_c/\omega_0 \propto \delta_0^{-2}$ is indicated in gray.}
\end{figure}

We plot the resonance frequency dependence with respect to $\mathcal{U}_0/k_B T$ and $\delta_0$ in Figs.~\ref{fig:TimeScaleU0} and~\ref{fig:TimeScaled0}. In Fig.~\ref{fig:TimeScaleU0} the linear dependence on $\mathcal{U}_0/k_B T$ expected from Eq.~(\ref{eq:omegac}) is clearly observed for intermediate values of $\mathcal{U}_0/k_B T$. For large values of $\mathcal{U}_0/k_B T$, we may observe the expected saturation when $\mathcal{U}_0/k_B T \simeq \delta_0^{-2}$ (in particular for $\mathcal{U}_0/k_B T \gtrsim 10$ and $\delta_0 = 0.2$; larger values of $\mathcal{U}_0/k_B T$ were not accessible either numerically or with the analytic expansion due to convergence issues.). For small values of the barrier strength $\mathcal{U}_0/k_B T$ the process becomes very weak and the scaling laws are no longer relevant. 

In Fig.~\ref{fig:TimeScaled0} the inverse quadratic dependence on $\delta_0$ is observed in a  narrow region, since it is expected for large $\mathcal{U}_0/k_B T$ and large $\delta_0$ (visible still for $\delta_0 \gtrsim 0.05$ and $\mathcal{U}_0/k_BT = 10$). For small values of $\delta_0$, the dependence of $\omega_c$ on $\delta_0$ is expected to saturate from Eq.~(\ref{eq:omegac}), and this is clearly observable in Fig.~\ref{fig:TimeScaled0}. In the intermediate regimes, more entangled dynamics are involved \mod{that may in particular require the introduction of other relevant time scales for the system}. We leave investigation of these more complex dynamics for future work.

Eq.~(\ref{eq:omegac}) provides a simplistic understanding of the dynamics involved and demonstrates that active osmotic flow may be strongly impacted by the specificities of the membrane in terms of asymmetry and solute interaction strength. Note that  the amplitude of the resonance may also be tuned with the different parameters at hand. As a rule of thumb, the greater the asymmetry (so for large values of the potential strength $\mathcal{U}_0/k_B T$ or small values of $\delta_0$), the greater the resonance.

\section{Energetic efficiency of active osmotic pumping}

\begin{figure}[h!]
\includegraphics*[width=0.99\columnwidth]{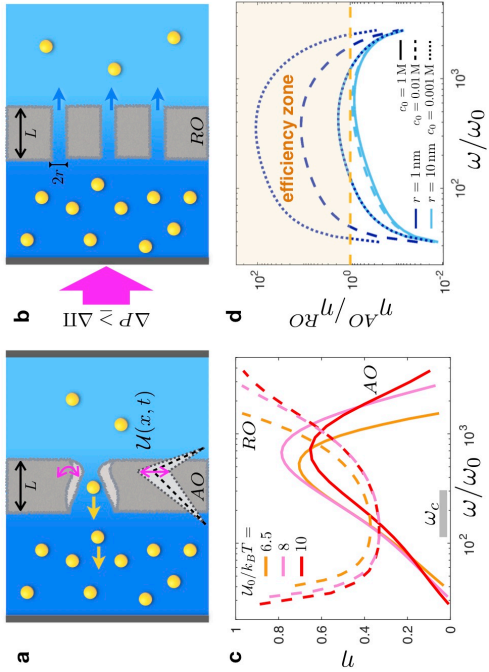}%
\caption{\label{fig:efficiency} \mod{\textbf{Efficiency of active osmosis versus reverse osmosis.} (a) Active osmosis with oscillating asymmetric barrier and fixed reservoir volumes. (b) Reverse osmosis counterpart, where a large external pressure is applied on one reservoir, driving solvent flow through pores impermeable to the solute. (c) Efficiency of both processes under the conditions where the thermodynamic collected energy  is the same in both cases (solid lines, active osmotic pumping as defined by Eq.~(\ref{eq:etaAO}) and calculated from simulations with $\delta_0 = 0.9$, $\epsilon = 1$ in the pump geometry with $C_1 = 1.82 c_0$ and $C_2 = 0.18 c_0$; dashed lines, reverse osmosis as defined by Eq.~(\ref{eq:etaRO}), with $r = 10 \mathrm{nm}$ and $c_0 = 0.001 \mathrm{M}$). The corresponding resonant osmotic frequency for the range of parameters used is indicated with a gray bar. (d) Efficiency of active osmosis as compared to reverse osmosis, as calculated from simulations with same parameters as in (c), $\mathcal{U}_0/k_BT = 8$ and molecular size $a = 1 \rm{\AA}$; for different values of $r$ and $c_0$ (here translated in mol/L). The efficiency zone corresponds to active osmotic pumping being more efficient than reverse osmosis.}  }
\end{figure}

%may be written as parrondo1998efficiency
In the context of filtration it is of utmost relevance to quantify the efficiency of the active osmotic process, and eventually compare it to other more common filtration processes. We consider the active osmosis (AO) configuration in a geometry similar to Fig.~\ref{fig:efficiency}-a, where the lateral reservoirs are closed and therefore the fluid flow $Q= 0$. When the membrane is dynamically activated -- \textit{e.g.} when the barrier $\mathcal{U}(x,t)$ is oscillated -- the average power spent writes (fully dimensionalized)~\cite{sekimoto1997kinetic}
\begin{equation}
\mathcal{P}^{AO} = \frac{1}{T} \int_0^T dt \int_{-L\delta_0}^{L\delta_1} S dx \, c(x,t) \frac{\partial \mathcal{U}(x,t)}{\partial t} 
\end{equation}
where $T = 2\pi /\omega$ and $S$ is the total accessible surface where the potential is exerted on the solute. The useful power generated by active osmosis corresponds to the chemical potential change of solute driven from one side to the other, which writes
\begin{equation}
\mathcal{P}_{\rm u}^{AO} = \langle J \rangle S k_B T \ln \frac{C_2}{C_1}.
\label{eq:PuAO}
\end{equation}
Therefore the efficiency of the active osmotic process is simply
\begin{equation}
\eta^{AO} = \frac{\mathcal{P}_{\rm u}^{AO}}{\mathcal{P}^{AO}}.
\label{eq:etaAO}
\end{equation}

We show in Fig.~\ref{fig:efficiency}-c the efficiency of the active osmotic process \mod{as a function of} the oscillation frequency $\omega$, for a set of parameters, varying only the membrane interaction strength $\mathcal{U}_0$. We find that the efficiency reaches a maximum (here up to $\eta^{AO} \simeq 0.8$) for a given value of the frequency, say $\omega_{\eta}$. Remarkably, $\omega_{\eta}$ is significantly higher than the resonance frequency $\omega_c$. In fact although the energy recovered $\mathcal{P}_{\rm u}^{AO}$ is indeed maximal for $\omega = \omega_c$, the energy expense $\mathcal{P}^{AO}$ is monotonically decreasing with $\omega$. This can be understood from the fact that at large frequencies solute has less time to diffuse around and therefore the energy expense to drive solute from a point to another is smaller. Furthermore, the maximal efficiency $\eta^{AO}(\omega_{\eta})$ strongly depends on the parameters of the system ($\delta_0$, $\Delta c$, $\mathcal{U}_0$). Although we do not carry here an in-depth study of these dependencies, we simply note that typically there is an optimal value for the membrane interaction strength $\mathcal{U}_0$. When $\UU_0 \ll k_BT$ there is almost no pumping flux; conversely, when $\UU_0 \gg k_B T$ more energy than needed is spent to drive the solute.

We now compare the active osmotic process to a prototypical filtration process: reverse osmosis, depicted in Fig.~\ref{fig:efficiency}-b. The reverse osmosis process similarly consists of two fluid reservoirs containing solvent and solute in concentration $C_1 > C_2$. The reservoirs are separated by a membrane which is permeable to the solvent alone (equivalent to a very large static barrier $\mathcal{U}(x,t)$, with $\mathcal{U}_0 \gtrsim 10 k_B T$). An operator applies a pressure in order to impose a reverse osmosis flow rate $Q$. The useful power extracted from the process corresponds to the reduction in mixing entropy of the system and writes
\begin{equation}
\mathcal{P}_{\rm u}^{RO} = Q (C_1-C_2) k_B T. 
\end{equation}
\mod{Note that this expression is not the same as for the AO process eq.~\ref{eq:PuAO}, which only involves transport of solute and no flow of solvent.} To compute the thermodynamic efficiency we now need to estimate the power that is dissipated. Without yet considering any physical membrane, the system necessarily dissipates energy through the friction of the solvent on the solute. Indeed, as solvent passes from the left reservoir to the right, it leaves behind the solute it contains, which gives rise to a relative velocity between the solvent and the solute particles. If we denote $L$ the characteristic thickness of the membrane and $S$ its surface area, % then during a time $\d t$, the solvent in the volume $L\cdot S$ moves at a velocity $Q/S$ by an amount $\d t Q/S$, with respect to $C_1 L S$ immobile solute particles. 
then each solute particle generates on the solvent a friction force equal to $\mu Q/S$, where $\mu = \frac{D}{k_B T}$ is the mobility of the solute. Since there are $C_1 L S$ immobile solute particles, \mod{and the solvent moves} with speed $Q S$, the power dissipated through friction is 
\begin{equation}
\mathcal{P}_{\rm f}^{RO} = \frac{C_1 L}{\mu S} Q^2
\end{equation}
If we now assume that the solvent has to pass through $n$ physical channels of circular cross-section area $s = \pi r^2$ (we assume $S = n s$), then we have to take into account the power dissipated through the hydrodynamic resistance of the channels, $R_h = 8 \pi \eta L/ s^2$, where $\eta$ is the viscosity of the solvent (assuming a no-slip boundary condition at the walls). The dissipated power reads 
\begin{equation}
\mathcal{P}_{\rm h}^{RO} = n R_h (Q/n)^2 = 8 \pi \eta L \frac{Q^2}{ns^2}.
\end{equation}
\mod{We have in fact estimated the hydrodynamic permeability $\mathcal{L}_{\rm hyd}$ of the RO membrane: 
\begin{equation}
\mathcal{P}_{\rm f}^{RO} + \mathcal{P}_{\rm h}^{RO} \equiv \frac{1}{\mathcal{L}_{\rm hyd}} \frac{Q^2}{S},
\end{equation}
with 
\begin{equation}
\mathcal{L}_{\rm hyd}^{-1} = \frac{s}{8 \pi \eta L} + \frac{C_1 L}{\mu}.
\end{equation}
Although this result relies on a model of discrete pores, it yields an estimate which agrees very well with the values reported for state-of-the-art RO polymeric membranes~\cite{Werber2016}, when evaluated for nanometre-sized pores. }

We may now compute the thermodynamic efficiency of the reverse osmosis process as 
\begin{equation}
\eta^{RO} = \frac{\mathcal{P}^{RO}_{\rm u}}{\mathcal{P}^{RO}_{\rm u} + \mathcal{P}^{RO}_{\rm f} + \mathcal{P}^{RO}_{\rm h}}
\end{equation}
and expanding
\begin{equation}
\eta^{RO} = \frac{1}{1+\dfrac{L}{D}\dfrac{C_1}{C_1-C_2}\dfrac{Q}{S} + \dfrac{8 \pi \eta}{(C_1-C_2)k_B T}\dfrac{L}{s} \dfrac{Q}{S}}.
\label{eq:etaRO1}
\end{equation}
\mod{As expected, the efficiency equals 1 for vanishing flow rate $Q$; however, it decreases at increasing flow rates.}

\mod{To compare the two processes, we require that they generate the same useful power. For a given AO current $\langle J \rangle$, this sets the RO flow rate $Q$ as $S\langle J \rangle \ln (C_2/C_1)/(C_2-C_1)$. 
Substituting in eq.~\eqref{eq:etaRO1} yields:}
\begin{equation}
\eta^{RO} =  \frac{1}{1+\dfrac{L \langle J \rangle}{D c_0}  \left(\dfrac{c_0}{\Delta C}\right)^2 \ln \dfrac{C_2}{C_1} \left( \dfrac{C_1}{c_0}+ \dfrac{4}{3\pi a r^2 c_0}\right)}
\label{eq:etaRO}
\end{equation}
where we made use of Einstein's relation $D = k_B T/ 6\pi \eta a$ with $a$ the molecular size of the solute. \mod{From Eq.~(\ref{eq:etaRO}), it is clear that RO becomes inefficient in the limit of very small pore sizes, where the hydrodynamic resistance is significant. Interestingly, it also shows that the efficiency is a decreasing function of $\langle J \rangle$, while the efficiency of AO is maximal around the highest values of $\langle J \rangle$. Therefore, we expect RO to be inefficient at the fluxes where AO is at its peak efficiency. This can be seen in particular in Fig.~\ref{fig:efficiency}-c where we show the efficiency of both processes.} 

We compare in Fig.~\ref{fig:efficiency}-d the efficiency of the reverse osmosis $\eta^{RO}$ and the active osmosis $\eta^{AO}$ processes for the optimal value of $\mathcal{U}_0$ at different forcing frequencies $\omega$. The results indeed show that there exists a broad range of parameters (for example nearly all concentrations $c_0 \lesssim 1$ M and $r = 1$ nm) where the active osmotic process is more efficient (and even up to 100 times more efficient) than the reverse osmosis process. This is extremely encouraging for filtration applications with \textit{active} membranes. Furthermore, from a more fundamental point of view, it is fascinating to see how it is possible to bypass the limitations of  filtration across \textit{static membranes} by injecting energy at the scale of membrane pores (and not at a macroscopic scale as is the case with reverse osmosis). %but at more local scale (the scale of the membrane). 
To some extent this echoes the ``apparent second principle breaking" in active matter (with active particles, self-spinners and so on~\cite{palacci2013living,aubret2018targeted}), where energy is also being consumed at the very local scale.
In this \textit{strongly out-of-equilibrium} regime, the principles underlying osmosis and selectivity can bypass the simple 'trade-off' picture of separation and has therefore a  great potential for new separation methodologies. 

%When $\mathcal{U}_0 \ll k_B T$, pumping is not happening in a very important amount. When $\mathcal{U}_0 \gg k_B T$, the energetic expense to drive solute down the barrier is in excess compared to that needed.

%

\section{Conclusion}

To summarize, we draw here a first picture to understand osmosis across active membranes, or \textit{out-of-equilibrium} osmosis. We provide a robust model to describe and account for the osmotic pressure as a function of the typical oscillating frequency of the membrane dynamics. Remarkably,  this kinetic model shows that osmotic flow through the membrane is still described by the Kedem-Katchalsky transport equations as~\cite{kedem1961physical,Kedem}
\begin{equation}
\langle Q \rangle = -\mathcal{L}_\text{hyd} \left( \Delta p - \sigma_{\rm app} k_B T \Delta C\right), \label{kk1}
\end{equation}
where $\sigma_{\rm app}$ is an apparent rejection coefficient that takes into account the specifics of the membrane and its dynamics. The solute flow (neglecting convection) may also be written
\begin{equation}
\langle J\rangle = - \frac{D}{L} \omega_{\rm app} \Delta C\label{kk2}
\end{equation}
where $\omega_{\rm app}$ still verifies the fundamental reciprocal relation $\omega_{\rm app} = 1 - \sigma_{\rm app}$. However all coefficients are now complex functions of the frequency of the active membrane. 

Our model clarifies the underlying principles of active osmosis. In particular, we have rationalized that at very low frequencies a dynamic membrane (\textit{e.g.} pore opening and closing) behaves as an apparently \textit{more permeable} membrane; whereas at very large frequencies a dynamic membrane behaves as an apparently \textit{static} membrane. In the intermediate regime, very interesting functionalities may be achieved, provided the membrane has some asymmetry: resonant pumping or sink, with a variety of tuneable features. Interestingly, active osmosis may be easily connected to potential ratchets and intuition from this field may be translated to the description of active osmosis. Finally, we demonstrate that in nanofiltration processes active osmosis may outperform reverse osmosis in terms of energetic efficiency. 

The model considered here is simple \mod{and provides a basis to study a number of effects. For example we expect (see Fig.~\ref{fig:Alternative2} in Appendix C) that asymmetry not just in space but also in time, \textit{e.g.} how fast the barrier is activated up versus down, may lead to more interesting regimes. Going further,} a number of details at the nanoscale could be accounted for, so as to provide a more systematic and thorough description of nanofiltration across membranes: this includes, for instance, electrostatic effects or surface interactions. \mod{The impact of noise (of the membrane interaction potential~\cite{marbach2018transport}, or due to the small number of solutes in the channel~\cite{smeets2008noise,Secchi2016}) on osmotic pressure is expected to be relevant at these scales and has to be explored.}  %?~\cite{gang1990periodically})
Such extensions will be the subject of future work. However the main generic features of active osmosis are expected to be captured by the present model.

Overall, our model, even simplistic, provides a number of rules of thumb to design active membrane, {\it e.g.} in terms of the asymmetry of the membrane or the typical frequency range at play. In practice composite membranes with tuneable sieving properties, for example gated by applied voltage, are a natural lead to explore the fabrication of such active membranes. 
Active osmosis through dynamic membranes has a considerable potential to broaden the current paradigm of filtration, building the basis for advanced filtration devices and artificial ionic machinery.

\section*{Acknowledgements}
%LB acknowledges support from the European Union's H2020 Framework Programme/FET Nanophlow. 
%LB acknowledges support from ERC, project Shadoks.
L.B. acknowledges funding from the EU H2020 Framework Programme/ERC Advanced Grant agreement number 785911-Shadoks.

\section*{Appendix}

\subsection*{Appendix A : Explicit solution of the triangular profile barrier}

\subsubsection*{Triangular profile} 

We assume that the potential is piece-wise linear, {\it i.e.}
 \begin{align}
 \phi(x)&=1+{x \over \delta_0}\,\, {\rm for}\,\, -\delta_0<x<0\nonumber \\
 \phi(x)&=1-{x\over \delta_1}\,\, {\rm for}\,\, 0<x<\delta_1\nonumber 
\end{align}
such that the force $\gamma = -\partial_x\phi = -1/\delta_0$ (resp. $+1/\delta_1$) for $x<0$ (resp. $x>0$).
Eq.~(\ref{Smolu4}) reduces to 
\begin{align}
j\omega \delta c_1(x)=& \partial_{xx} \delta c_1(x)   - {u_0}  \gamma \partial_x \delta c_1(x)  \, -\gamma \partial_x f(x) ,
\label{Smolu5}
\end{align}
where we introduced $f(x)=  \,{u_0}\cc(x)$.
The average concentration $\cc(x)$ is easily computed as, for $x <0$
\begin{align}
\cc(x) =(1 - \Delta c) e^{-u_0(1+x/\delta_0)} +  \Delta c \bigg[ \delta_0 {e^{-u_0(1 + x/\delta_0)} - 1\over (e^{u_0}-1)} \bigg]
\end{align}
and for $x>0$:
\begin{equation}
\cc (x) = e^{-u_0(1-x/\delta_1)} - \Delta c \left[  \delta_1 {e^{-u_0(1-x/\delta_1)} - 1\over (e^{u_0}-1)} \right]
\end{equation}

Eq.~(\ref{Smolu5})e and a full expansion at second order for $c(x)$ can be readily calculated. The osmotic pressure can be deduced accordingly. On the left domain or $x <0$, Eq.~(\ref{Smolu5}) can be rewritten
\begin{equation}
\partial_{xx} \delta c_1(x)   + {u_0 \over \delta_0} \partial_x \delta c_1(x)  - j\omega \delta c_1(x) =  - {1 \over \delta_0 }\partial_x f(x) ,
\end{equation}
and on the right domain:
\begin{equation}
\partial_{xx} \delta c_1(x)   - {u_0 \over \delta_1} \partial_x \delta c_1(x)  - j\omega \delta c_1(x) =  {1 \over \delta_1 }\partial_x f(x).
\end{equation}

\subsubsection*{Expression of $\delta c_1$}
Let us introduce 
\begin{equation}
\lambda_\pm^L= {1\over 2} \left( \textcolor{blue}{-}{u_0\over \delta_0} \pm \sqrt{ \left({u_0 \over \delta_0}\right)^2+4j\omega}\right)
\end{equation}

\begin{equation}
\lambda_\pm^R= {1\over 2} \left({u_0\over \delta_1} \pm \sqrt{ \left({u_0\over \delta_1}\right)^2+4j\omega}\right)
\end{equation}

The solution for Eq.(\ref{Smolu5}) then writes, for $x <0$
\begin{align}
\delta c_1(x)=& \alpha_L e^{\lambdamL x} + \beta_L e^{\lambdapL x}\nonumber \\
&+ {e^{\lambdamL x}\over \sqrt{ \left({u_0\over \delta_0}\right)^2+4j\omega}} \int_0^x dx^\prime \, e^{-\lambdamL x^\prime} (\partial_x f)(x^\prime) \nonumber \\
&- {e^{\lambdapL x}\over \sqrt{ \left({u_0 \over \delta_0}\right)^2+4j\omega}} \int_0^x dx^\prime \, e^{-\lambdapL x^\prime} (\partial_x f)(x^\prime) 
\end{align}
and for $x> 0$
\begin{align}
\delta c_1(x)=& \alpha_R e^{\lambdamR x} + \beta_R e^{\lambdapR x}\nonumber \\
&- {e^{\lambdamR x}\over \sqrt{ \left({u_0\over \delta_1}\right)^2+4j\omega}} \int_0^x dx^\prime \, e^{-\lambdamR x^\prime} (\partial_x f)(x^\prime)\nonumber \\
&+ {e^{\lambdapR x}\over \sqrt{ \left({u_0\over \delta_1}\right)^2+4j\omega}} \int_0^x dx^\prime \, e^{-\lambdapR x^\prime} (\partial_x f)(x^\prime).
\end{align}

\subsubsection*{Boundary conditions}
The boundary conditions are $\delta c(x=-\delta_0)=\delta c(x=\delta_1)=0$. This imposes
\begin{align}
0=& \alpha_L e^{-\lambdamL\delta_0} + \beta_L e^{-\lambdapL\delta_0}\nonumber \\
&+ {e^{-\lambdamL\delta_0}\over \sqrt{ \left({u_0 \over \delta_0}\right)^2+4j\omega}} \int_0^{-\delta_0} dx^\prime \, e^{-\lambdamL x^\prime} (\partial_x f)(x^\prime) \nonumber \\
&- {e^{-\lambdapL\delta_0}\over \sqrt{ \left({u_0 \over \delta_0}\right)^2+4j\omega}} \int_0^{-\delta_0} dx^\prime \, e^{-\lambdapL x^\prime} (\partial_x f)(x^\prime) 
\label{CL1}
\end{align}
and
\begin{align}
0=& \alpha_R e^{\lambda_-^R \delta_1} + \beta_R e^{\lambda_+^R \delta_1}\nonumber \\
&- {e^{\lambda_-^R \delta_1}\over \sqrt{ \left({u_0\over \delta_1}\right)^2+4j\omega}} \int_0^{\delta_1} dx^\prime \, e^{-\lambda_-^R x^\prime} (\partial_x f)(x^\prime)\nonumber \\
&+ {e^{\lambda_+^R\delta_1}\over \sqrt{ \left({u_0\over \delta_1}\right)^2+4j\omega}} \int_0^{\delta_1} dx^\prime \, e^{-\lambda_+^R x^\prime} (\partial_x f)(x^\prime)
\label{CL2}
\end{align}
with $f(x)= +  \,{u_0}\cc(x)$.

Two more subtle conditions are continuity conditions at $x=0$.
The continuity of the concentration imposes $\delta c_1(0^-)=\delta c_1(0^+)$, so that
\begin{equation}
\alpha_R+\beta_R=\alpha_L+\beta_L
\label{CL3}
\end{equation}
The condition for the continuity of the (first order) flux can be obtained by integrating Eq.~(\ref{Smolu5}) between $x=0^-$ and $x=0^+$,
which imposes
\begin{align}
&\partial_x \delta c_1(0^+)-{u_0 \over \delta_0} \delta c_1(0^+)- {u_0 \over \delta_0} \cc(0)\nonumber \\
&=\partial_x \delta c_1(0^-)+{u_0\over \delta_1} \delta c_1(0^-)+ {u_0\over \delta_1} \cc(0)
\end{align}
After calculating the terms $\partial_x \delta c_1(0^-)=\lambdamL\alpha_L+\lambdapL\beta_L$ and $\partial_x \delta c_1(0^+)=\lambdamR \alpha_R +\lambdapR\beta_R$, one deduces the continuity equation
\begin{align}
-\lambdapL\alpha_L+\lambdapR \alpha_R-\lambdamL \beta_L+\lambdamR\beta_R = - u_0 \left[{1\over \delta_0\delta_1}\right]\,\cc(0) 
\label{CL4}
\end{align}
where we used the expressions of the $\lambda$'s to simplify things.

\subsubsection*{Full Solution}

The system of Eqs. (\ref{CL1}), (\ref{CL2}), (\ref{CL3}), (\ref{CL4}) can be solved to obtain the explicit expressions for
$\alpha_\pm(\omega)$, $\beta_\pm(\omega)$ as a function of frequency and potential parameters. We don't provide the full expressions here, since they are highly cumbersome. We investigate the results in the main text on several limiting situations. 

\subsection*{Appendix B : Numerical simulation details}

The Smoluchowski equations are solved with a finite difference scheme over 6 orders of magnitude of $\omega/\omega_0$ and various other parameters. To ensure global convergence, we perform a Crank-Nicholson scheme and are especially careful that advection only carries upstream solute. The time step and space discretization were chosen such that any reduction of either one (\textit{e.g.} by a factor 2) leads to no significant numerical difference in the results. The initial concentration profile corresponds to the static barrier for $\epsilon = 0$. As we seek averages over the oscillating process, we look for the average of the osmotic pressure over five periods. When the simulation of an extra period will not change the osmotic pressure by a significant amount, the initial conditions are forgotten and the result is converged.

In the simulations time is nondimensionalized by $\omega$ such that typical simulations will roughly take the same time to run. Note that for very large frequencies, the relaxation from the initial conditions is much slower as the allowed flux is much smaller, and therefore simulations where run for longer times in the that case. 

The critical frequency at which the process is resonant corresponds to the frequency at which the osmotic reflection coefficient is maximum or minimum. As the simulation provides the osmotic reflection coefficient at only discrete values of the frequency $\omega$ we perform a fit on a very narrow region around the maximum (resp. minimum; with a 4th order standard polynomial fit to account for peak slight distortion) and obtain the critical frequency from this fit. For each fit the agreement with the simulation data is thoroughly asserted such that the critical frequency obtained is a reliable value. 

\subsection*{Appendix C : Toy model for the asymmetric potential profile}

We consider a time-dependent triangular potential, with a spatial extension similar to  
the previous analysis, {\it i.e.}
 \begin{align}
 \phi(x)&=1+{x\over \delta_0}\,\, {\rm for}\,\, -\delta_0<x<0\nonumber \\
 \phi(x)&=1-{x\over \delta_1}\,\, {\rm for}\,\, 0<x<\delta_1  \\
\end{align}
where $x$ is the dimensionless coordinate (in units of the membrane width, say $L$); $\delta_0$, $\delta_1$ are
in dimension of $L$ ($\delta_0+\delta_1=1$).

But we now consider a simplified time-dependence, where this triangular potential is periodically 
ON/OFF for time-lapse with period $T$:
\begin{equation}
{\cal U}(x,t)={\cal U}_0\times f(t)\times \phi(x)
\end{equation}
with $f(t)=0$ for $t\in[k T; k(T+\tau_1)]$ and $f(t)=1$ for $t\in[ k(T+\tau_1); (k+1)T]$, with $k=E(t/T)$, an integer.
Note that $t$ is here in units of $\tau_0=L^2/D$ the diffusion time-scale.

Boundary conditions for the concentration in the reservoirs are: $C_1$ for $x<-\delta_0$  and $C_2$ for $x>\delta_1$.

We will make several simplifying assumptions to obtain work out the model and obtain tractable results. 
First we assume that at the ON period, with duration $\tau_2=T-\tau_1$ is sufficiently long so that 
particles reach an equilibrium state in the potential. This ``re-initializes'' the problem after each period $T$.
Second we wil assume that the energy barrier ${\cal U}_0$
is very large, so that no particle can cross when the potential is on. Also such a high potential will basically confine particles for $x<-\delta_0$ and $x>\delta_1$; {\it i.e.} we neglect the extension of the equilibrium density profile in the region $[-\delta_0;\delta_1]$ when the potential is ON.

Under these simplified assumptions, some interesting predictions can be obtained. 
We recall that the solution for free diffusion with initial condition $c(x,t=0)=\Theta(x)$ (Heaviside) and 
boundary conditions $c(x=0,t)=1$, $c(x\rightarrow \infty,t)=0$ is
\begin{equation}
c(x,t)=\psi\left[ {x \over 2\sqrt{t}}\right]
\end{equation}
with 
\begin{equation}
\psi(x)=\left[1-{1\over 2}\left(1+{\rm Erf} (x) \right)
+{1\over 2}\left(1+{\rm Erf} (-x)\right)
 \right] 
\end{equation}

Accordingly, once the potential is released (ON$\rightarrow$ OFF period), one may simplify the solution for the concentration by superposing diffusion from the two reservoirs into the membrane: 
\begin{equation}
C(x,t)=C_1 \times \psi\left[ {x+\delta_0 \over 2\sqrt{ t}}\right] + C_2 \times 
\psi\left[ {\delta_1-x \over 2\sqrt{t}}\right]
\end{equation}

The flux is defined as the number 
of particles which cross the barrier maximum at $x=0$ in the OFF period. Indeed, once the potential is back to ON, the particle for $x>0$ will be carried on to the right, while the particles for $x<0$ will be carried on to the left.
Then the flux is accordingly defined as
\begin{equation}
\langle J \rangle={1\over T} \times (N_R-N_L)
\end{equation}
with
\begin{align} 
&N_R=\int_0^{\delta_1}dx\, C_1 \times \psi\left[ {x+\delta_0 \over 2\sqrt{ \tau_1}}\right] \nonumber \\
&N_L=\int_{-\delta_0}^0 dx\, C_2 \times \psi\left[ {\delta_1-x \over 2\sqrt{ \tau_1}}\right] 
\end{align}
{\it i.e.} the number of particles which have crossed $x=0$ (from left to right, or right to left) at the time $\tau_1$.

Let us introduce $\Psi(x)=\int_0^x \psi(x) dx$. One may calculate:
\begin{equation}
\Psi(x)={1-e^{-x^2}\over \sqrt{\pi}}+x \, (1- {\rm Erf} (x) )
\end{equation}

\begin{figure}[h!]
\includegraphics*[width=1.\columnwidth]{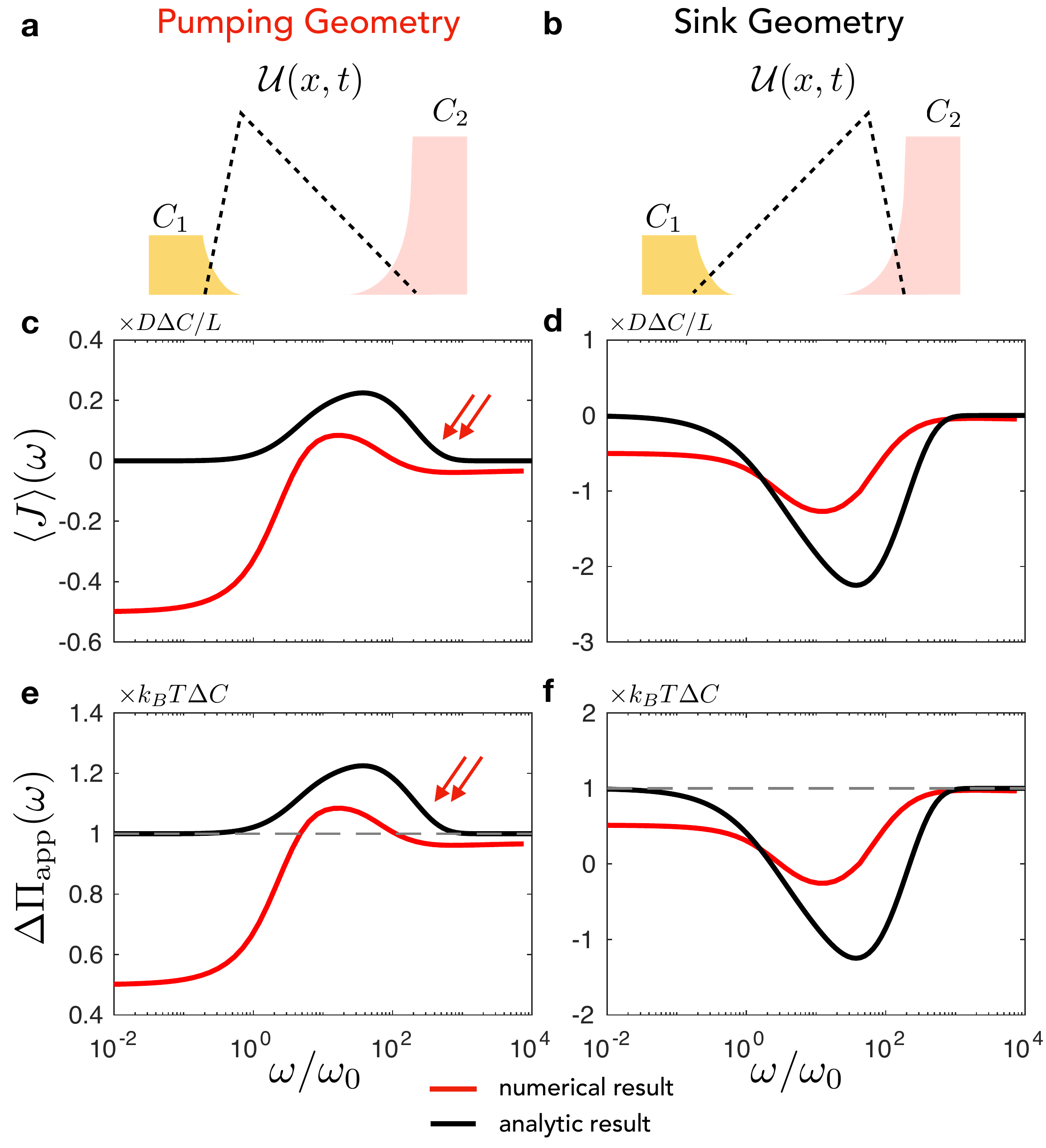}%
\caption{\label{fig:Alternative} {\bf Toy model of resonant osmosis} (a) Schematic showing the geometry relevant for pumping, with a steep energy barrier near the low concentration, $C_2 > C_1$ and $\delta_0 < 0.5$. (b) Schematic showing the opposite geometry $C_2 > C_1$ but $\delta_0 > 0.5$. (c) and (d) Simulated (orange lines) and analytic (black lines) results for the effective normalized flux $\langle J \rangle (\omega)$ Eq.~\ref{eq:J2} and (e) and (f) apparent osmotic pressure $\Delta\Pi_{\rm app}(\omega)$ Eq.~\ref{Rejection2}. The results are plotted for $\delta_0 = 0.1$ in the pumping geometry and $\delta_0 = 0.9$ in the sink geometry, $C_1=0.1$, $C_2=1.0$; and in the numerical computation $\UU_0 = 10 k_B T$. The couple of red arrows indicates regimes where pumping is seen, \mod{and the ON-OFF times are kept equal $\alpha = 0.5$.} }
\end{figure}

Then the flux (in units of $D/L$) is
\begin{align}
\langle J \rangle={1\over T} \times &2\sqrt{ \tau_1}\times \biggl[
C_1\times \left(\Psi\left({1\over2\sqrt{ \tau_1}}\right)-\Psi\left({\delta_0\over2\sqrt{ \tau_1}}\right)\right)
 \nonumber \\
&
-C_2
\times \left(\Psi\left({1\over2\sqrt{ \tau_1}}\right)-\Psi\left({\delta_1\over2\sqrt{ \tau_1}}\right)\right)\biggr] 
\end{align}
The characteristic frequency is $\omega_0=2\pi  /\tau_0$. Then $T=2\pi/\omega$ and $\tau_1=\alpha\times 2\pi/\omega$.

Now we rewrite the expression in terms of frequency, $\omega=2\pi/T$. Writing $\tau_1=\alpha T$ (with $\alpha$ the fraction of time with OFF potential), one obtains the flux in units of $L/D$ as
 \begin{align}
\langle J \rangle {L\over D}=&2\sqrt{{ \alpha\, \omega \over \omega_0}}\times \biggl[ ...
\nonumber \\
&C_1 \left(\Psi\left({1\over2\sqrt{ \alpha }}\sqrt{{\omega\over \omega_0}}\right)-\Psi\left({\delta_0\over2\sqrt{ \alpha }}\sqrt{{\omega\over \omega_0}}\right)\right)
 \nonumber \\
&
-C_2 \left(\Psi\left({1\over2\sqrt{ \alpha }}\sqrt{{\omega\over \omega_0}}\right)-\Psi\left({\delta_1\over2\sqrt{ \alpha }}\sqrt{{\omega\over \omega_0}}\right)\right)\biggr] 
\label{eq:J2}
\end{align}

The osmotic pressure is accordingly defined as
\begin{equation}
  \Delta\Pi= k_BT\,[C_2-C_1] +{k_BT\over D}\times {L} \times \langle J \rangle
 \end{equation}
 (with $L=\delta_0+\delta_1$)
 and the apparent rejection coefficient is
   \begin{align}
 \sigma_{\rm app}&[\omega,C_1,C_2]=1+{L\over D} \times {\langle J \rangle\over C_2-C_1}
   \label{Rejection2}
  \end{align}

The frequency dependent flux and osmotic rejection coefficient are plotted in Fig.~\ref{fig:Alternative}, with several interesting features. First a resonance is clearly observed. What is remarkable is that (i) a pump behavior is observed (left panels, : $J>0$ while $C_2>C_1$) and (ii) a change of sign is observed for the osmotic rejection coefficient. The latter means that one can tune the sign of the osmotic pressure and it can even be made vanish for a given frequency!

\begin{figure}[h!]
\includegraphics*[width=1.\columnwidth]{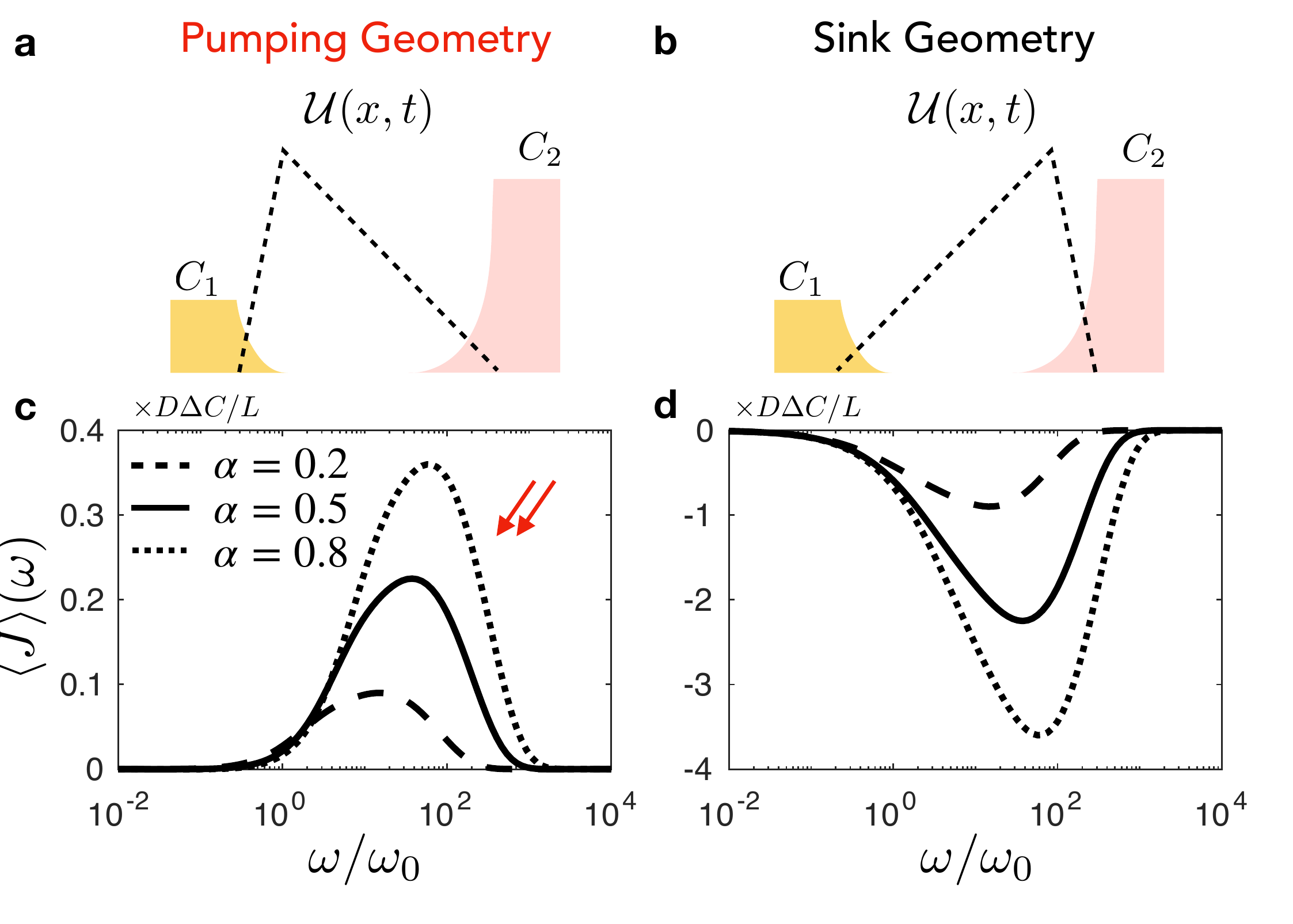}%
\caption{\label{fig:Alternative2} \mod{ {\bf Asymmetric ON/OFF pumping} (a) Schematic showing the geometry relevant for pumping, with a steep energy barrier near the low concentration, $C_2 > C_1$ and $\delta_0 < 0.5$. (b) Schematic showing the opposite geometry $C_2 > C_1$ but $\delta_0 > 0.5$. (c) and (d) Analytic results for the effective normalized flux $\langle J \rangle (\omega)$ Eq.~\ref{eq:J2} for different ratios of the ON-OFF respective times of the barrier. Note that $\alpha$ measures how long the barrier is OFF. The results are plotted for $\delta_0 = 0.1$ in the pumping geometry and $\delta_0 = 0.9$ in the sink geometry, $C_1=0.1$, $C_2=1.0$; and in the numerical computation $\UU_0 = 10 k_B T$. The couple of red arrows indicates regimes where pumping is seen. } }
\end{figure}

\mod{Finally, note that the tow model is a very good proxy to build insight on the effect of asymmetric barriers not just in space but also \textit{in time}. In Fig.~\ref{fig:Alternative2} we show how asymmetry in time dramatically impacts solute flux around the resonance frequency. We observe that the longer the barrier is OFF, the more solute is pumped (or is sunk). This makes sense considering that the longer the barrier is OFF, the more solute can actually go past the the barrier peak. To improve our insight on these different regimes further computations have to be done that we leave for future work.}

\end{document}